\documentclass[eqsecnum,preprint,prd,aps,nofootinbib]{revtex4}
\usepackage{amssymb,amsmath,amsfonts}
\usepackage{bm}
\usepackage{boxedminipage}
\usepackage[pdftex]{graphicx}
\usepackage{graphicx}
\usepackage{enumerate}

\newcommand{\be}{\begin{equation}}
\newcommand{\ee}{\end{equation}}
\newcommand{\ba}{\begin{eqnarray}}
\newcommand{\ea}{\end{eqnarray}}
\newcommand{\bc}{\begin{center}}
\newcommand{\ec}{\end{center}}
\def\lvec#1{\vbox{\ialign{##\crcr$\leftarrow$\crcr\noalign{
 \kern-1pt\nointerlineskip}$\hfil\displaystyle{#1}\hfil$\crcr}}}
\def\rvec#1{\vbox{\ialign{##\crcr$\rightarrow$\crcr\noalign{
 \kern-1pt\nointerlineskip}$\hfil\displaystyle{#1}\hfil$\crcr}}}

\begin{document}
\begin{center}
\bibliographystyle{article}

 { \Huge{\textbf{\textbf Attractor Flows in {$st^{2}$} Black Holes }}}

\end{center}
\vspace{0.4cm}


\author{Raju Roychowdhury$^{1,2}$ 
\thanks{Electronic address: raju@na.infn.it}}

\affiliation{
${\ }^{1}$Istituto Nazionale di Fisica Nucleare, Sezione di Napoli,\\
Complesso Universitario di Monte S. Angelo, Via Cintia, Edificio 6, 80126
Napoli, Italy\\
${\ }^{2}$Dipartimento di Scienze Fisiche, Federico II University,\\
Complesso Universitario di Monte S. Angelo, Via Cintia, Edificio 6,
80126 Napoli, Italy}

\vspace{0.4cm}
\date{\today}

\begin{abstract}
Following the same treatment of Bellucci et.al., we obtain, the hitherto unknown general solutions of the radial attractor flow equations for extremal black holes, both for non-BPS with non-vanishing and vanishing central charge $Z$ for the\ so-called $st^2$ model, the minimal \textit{rank-}$2$ $\mathcal{N}=2$ symmetric supergravity in $d=4$ space-time dimensions.

We also make useful comparisons with results that already exist in literature, and introduce the \textit{fake supergravity (first-order) formalism} to be used in our analysis. An analysis of the
\textit{BPS bound} all along the non-BPS attractor flows and of the
\textit{marginal stability} of corresponding $D$-brane charge configurations has also been presented.

\end{abstract}

\maketitle
\bigskip
\vspace{2cm}

\section{Introduction}
Black Holes  \cite{witt}--\nocite
{moore,duff1,hawking1,penrose}\cite{gibbons1} are truely unique objects for theoretical physicists as they pose various fascinating problems, which may offer a clue for solving the riddle of Quantum Gravity. One of the recent developments in the arena of Black Hole Physics is the issue of the {\it Attractor Mechanism} \cite{ferrara1}--\nocite{ferrara2,strominger2}\cite{FGK}, a
remarkable phenomenon occurring in case of \textit{extremal} BHs coupled to Maxwell
and scalar fields, in supersymmetric theories of gravity
\cite{Sen-old1}--\nocite{GIJT,Sen-old2,K1,TT,G,GJMT,Ebra1,K2,Ira1,Tom,
BFM,AoB-book,FKlast,Ebra2,bellucci1,rotating-attr,K2-bis,Misra1,Lust2,Morales,BFMY,Astefa,CdWMa, DFT07-1,BFM-SIGRAV06,Cer-Dal-1,ADFT-2,Saraikin-Vafa-1,Ferrara-Marrani-1,TT2, ADOT-1,ferrara4,CCDOP,Misra2,Astefanesei,Anber,Myung1,Ceresole,BMOS-1,Hotta, Gao, PASCOS07,Sen-review,Belhaj1,AFMT1,Gaiotto1,BFMS1,GLS-1,ANYY1,bellucci2,Cai-Pang,Vaula, Li,BFMY2,Saidi2,Saidi3,Saidi4,Valencia-RTN-07,FHM,Unattractor,Vaula2,Trigiante}
\cite{Gnecchi-1} (for further developments, see also \textit{e.g.} \cite{OSV}%
--\nocite{OVV,ANV}\cite{GSV}).

Supergravity \cite{VN-Supergravity} is the low-energy limit of \textit{superstrings} 
\cite{maldacena}--\nocite{schwarz1,schwarz2} \cite{gasperini} or \textit{M-theory } 
\cite{witten,schwarz3,schwarz3-bis}; in such a framework, a certain number of 
abelian gauge fields and moduli fields are coupled to the
Einstein-Hilbert action. This is true for the theories in
$d=4$ space-time dimensions, and having $\mathcal{N}\geqslant 2$
\textit{supercharges}, where $4\mathcal{N}$ is the number of
supersymmetries. The fermionic sector of these theories contains a certain number of spin $%
1/2 $ fermions and $\mathcal{N}$ spin $3/2$ Rarita-Schwinger fields, i.e.
 the \textit{gravitinos} (the gauge fields of local supersymmetry). The vanishing
of the supersymmetric variation of the gravitinos determines whether or not
a certain number of supersymmetries (\textit{BPS property}) is preserved by
the BH background.

In this setting, asymptotically flat charged BH solutions, within a static
and spherically symmetric \textit{Ansatz}, mimic the famous Schwarzschild BH. 
A remarkable feature of electrically (and/or magnetically) charged BHs \cite{nordstrom}
as well as rotating ones \cite{smarr} is a somewhat unconventional
thermodynamical property called \textit{extremality} \cite
{gibbons1,kallosh1,SUSY-censor}. \textit{Extremal} BHs are possibly stable
gravitational objects with finite entropy but vanishing temperature. 
\textit{Extremality} also means that the inner
(Cauchy) and outer (event) horizons do coincide, thus implying vanishing
\textit{surface gravity} (for a recent review see \textit{e.g.} \cite{FHM},
and Refs. therein).

In the regime of \textit{extremality} a particular relation among entropy,
charges and spin holds, yielding that the Arnowitt-Deser-Misner (ADM)
mass \cite{arnowitt,deser,bondi} is \textit{not} an independent quantity.
A beautiful phenomenon happens for Black Hole physics as the
\textit{No Hair theorem} states that there is a limited number of parameters
which describe space and physical fields far away from the Black Hole.
In application to the recently studied Black Holes in String theory, the parameters 
include mass, electric and magnetic charges and the asymptotic values of the 
scalar fields. These values may continuously vary, being an
arbitrary point in the moduli space of the theory or, in a more geometrical
language, a point in the target manifold of the scalar non-linear Lagrangian
\cite{ferrara1,gibbons3}. It appears that for SUSY Black Holes one  can prove
a stronger version of the \textit{No Hair theorem}: Black Holes loose all their
scalar hair near the horizon and their solutions in the near horizon limit are characterized 
only by a discrete set of parameters which correspond to conserved charges associated 
with the gauge symmetries. Nevertheless, the BH entropy, as given by the
Bekenstein-Hawking entropy-area formula \cite{hawking2}, is independent of the scalar charges (\textit{``no scalar hair''}) and it only depends on the asymptotic (generally \textit{dyonic}) BH charges in this case.

All extremal static, spherically symmetric and asymptotically flat BHs in $%
d=4$ have a \textit{Bertotti-Robinson} \cite{bertotti} $AdS_{2}\times S^{2}$
near-horizon geometry, with vanishing scalar curvature and \textit{%
conformal flatness}; in particular, the radius of $AdS_{2}$ coincides with the
radius of $S^{2}$, and it is proportional to the (square root of the) BH
entropy (in turn proportional, through the Bekenstein-Hawking formula \cite
{hawking2}, to the area of the event horizon). Non-BPS (\textit{i.e.}
non-supersymmetric) (see \textit{e.g.} \cite
{FGK,bellucci1,TT2,ferrara4,Gaiotto1,GLS-1,ANYY1}) extremal BHs exist as
well, and they also exhibit an attractor behavior.\bigskip

A particularly remarkable model in $\mathcal{N}=2$, $d=4$ \textit{ungauged}
supergravity is the so-called $st^{2}$ model. It has
been recently shown to be relevant for the special entangled quantum systems
and the Freudenthal construction involving a three-qubit system consisting of 
one distinguished qubit and two bosonic qubits \cite{VLevay}.

The $2$ complex scalars coming from the two Abelian vector multiplets
coupled to the supergravity one span the \textit{rank-}$2$, completely
factorized special K\"{a}hler manifold $\frac{G}{H}=\left( \frac{SU\left(
1,1\right) }{U\left( 1\right) }\right) ^{2}$, with $dim_{\mathbb{C}}=2$,
\begin{equation}
G=\left( SU\left( 1,1\right) \right) ^{2}\sim \left( SO\left( 2,1\right)
\right) ^{2}\sim \left( SL\left( 2,\mathbb{R}\right) \right) ^{2}\sim \left(
Sp\left( 2,\mathbb{R}\right) \right) ^{2}  \label{GG}
\end{equation}
being the $d=4$ $U$-duality group\footnote{%
With a slight abuse of language, we refer to $U$-duality group as to the
\textit{continuous} version, valid for large values of charges, of the
string duality group introduced by Hull and Townsend \cite{Hull}.}, while $%
H=\left( U\left( 1\right) \right) ^{2}\sim \left( SO\left( 2\right) \right)
^{2}$ is its \textit{maximal compact subgroup}. Such a space is nothing but the
element $n=-1$ of the cubic sequence of reducible homogeneous symmetric special
K\"{a}hler manifolds $\frac{SU(1,1)}{U(1)}\otimes \frac{SO(2,2+n)}{%
SO(2)\otimes SO(2+n)}$ (see \textit{e.g.}~\cite{bellucci1} and Refs. therein).

The $st^{2}$ model has $1$ non-BPS $Z\neq 0$ \textit{flat} directions, spanning
the moduli space $SO\left( 1,1\right)$ (\textit{%
i.e.} the scalar manifold of the $st^{2}$ model in $d=5$), but \textit{no}
non-BPS $Z=0$ \textit{massless Hessian modes} at all (see
also \cite{TT2} and \cite{Ferrara-Marrani-1} for a similar treatment for $stu$ model ). 
In other words, the $4\times 4 $ Hessian matrix of the effective BH potential at its non-BPS $Z\neq 0$
critical points has $3$ strictly positive and $1$ vanishing eigenvalues
(the latter correspond to \textit{massless Hessian modes}), whereas at its
non-BPS $Z=0$ critical points all the eigenvalues are strictly positive.
After \cite{FGK}, $\frac{1}{2}$-BPS critical points of $V_{BH}$ in $\mathcal{%
N}=2$, $d=4$ supergravity are \textit{all} stable, and thus they determine
attractors in a strict sense. 

Concerning its stringy origin, the $st^{2}$ model, is obtained e.g. by a
$t=u$ degeneracy of the so called $stu$ model which can be interpreted as 
the low-energy limit of Type $IIA$ superstrings compactified on a
six-torus $T^{6}$ factorized as $T^{2}\times T^{2}\times T^{2}$. The $%
D0-D2-D4-D6$ branes wrapping the various $T^{2}$s determine the $3$ magnetic
and $3$ electric BH charges.

The present paper studies in detail, the attractor
flow equations of the $st^{2}$ model, whose fundamental facts are summarized in
Sect. \ref{st2}. In a nutshell, we reformulate all the computations done by Bellucci 
et. al. in \cite{stuunveiled} for $stu$ model in the case of the much less known $st^{2}$ model
filling the vast gap in the existing supergravity black hole literature. 
All the classes of \textit{non-degenerate} (\textit{i.e.}
with non-vanishing classical Bekenstein-Hawking \cite{hawking2} BH entropy)
attractor flow solutions of the $st^{2}$ configuration are determined, in their most general form (with all $B$\textit{-fields} switched on). The main results of our investigation
are listed below:
\begin{itemize}
\item  As mentioned above, the $\frac{1}{2}$-BPS attractor flow solution is
known since \cite{Cvetic-Youm}-\nocite
{Tseytlin,Gauntlett,Bala,BPS-flow-1,BPS-flow-2,BPS-flow-3,Bates-Denef}\cite
{bala-foam}, and it is reviewed in Sect. \ref{BPS-Flow}. In Sect. \ref
{Non-BPS-Z=0-Flow} the non-BPS $Z=0$ attractor flow solution, untreated so
far, for the $st^{2}$ case is determined for the most general supporting BH charge configuration,
and its relation to the supersymmetric flow, both at and away from the event
horizon radius $r_{H}$, is established.

\item  Sect. \ref{Non-BPS-Z<>0-Flow} is devoted to the study of the non-BPS $%
Z\neq 0$ attractor flow solution in full generality. By using suitable $U$%
-duality transformations (Subsect. \ref{U-Duality-Transf}), and starting
from the $D0-D6$ configuration (Subsect. \ref{D0-D6}), the non-BPS $Z\neq 0$
attractor flow supported by the most general $D0-D2-D4-D6$ configuration
(with \textit{all} charges switched on) is explicitly derived in Subsect.
\ref{D0-D2-D4-D6}. This completes and generalizes the analyses performed in
\cite{K2-bis}, \cite{Hotta}, \cite{GLS-1} and \cite{Cai-Pang} for 
the $stu$ model to the $st^{2}$ case. It is also confirmed that in such a general
framework: the moduli space $SO\left( 1,1\right)$, known
to exist at the non-BPS $Z\neq 0$ critical points of $V_{BH}$ \cite
{ferrara4,TT2}, is found to be present \textit{all along} the non-BPS
attractor flow, \textit{i.e.} for every $r\geqslant r_{H}$.

\item  In Sect. \ref{Analysis-Particular} a detailed analysis of particular
configurations, namely $D0-D4$ (\textit{magnetic}, Subsect. \ref{D0-D4}),
its \textit{dual} $D2-D6$ (\textit{electric}, Subsect. \ref{D2-D6}), and $%
D0-D2-D4$ (Subsect. \ref{D0-D2-D4}), is performed.

\item  The so-called \textit{first order (fake supergravity) formalism},
introduced in \cite{Fake-Refs}, has been recently developed in \cite
{Cer-Dal-1} and \cite{ADOT-1} in order to describe $d=4$ \textit{extremal}
BHs; in general, it is based on a suitably defined real, scalar-dependent,
\textit{fake superpotential}\footnote{%
It is worth pointing out that the \textit{first order formalism}, as
(re)formulated in \cite{Cer-Dal-1} and \cite{ADOT-1} for $d=4$ \textit{%
extremal} BHs, automatically selects the solutions which do \textit{not}
blow up at the BH event horizon. In other words, the \textit{(covariant)
scalar charges} $\Sigma _{i}$ built in terms of the \textit{fake
superpotential} $\mathcal{W}$ (see Eq. (\ref{d-2}) further below) satisfy by
construction all the conditions in order for the \textit{Attractor Mechanism}
to hold. It should be here recalled that for \textit{extremal} BHs the solution
converging at the BH event horizon ($r\rightarrow r_{H}^{+}$) does \textit{%
not} depend on the initial, asymptotical values of the scalar fields. See%
\textit{\ e.g.} discussions in \cite{Astefa} and \cite{Astefanesei}.}\textit{%
\ }$\mathcal{W}$. In the framework of $st^{2}$ model, we explicitly build up $%
\mathcal{W}$ in the non-trivial cases represented by the non-BPS attractor
flows. For the non-BPS $Z=0$ attractor flow (Sect. \ref{Non-BPS-Z=0-Flow}) and
for  non-BPS $Z\neq 0$ attractor flow (Subsect.\ref{D0-D2-D4-D6}) \textit{fake superpotential}
$\mathcal{W}$ is also determined. 

\item  Within the \textit{first-order (fake supergravity) formalism}, for
all attractor flows we compute the \textit{covariant scalar charges }as well
as the \textit{ADM mass}, studying the issue of \textit{marginal stability}
\cite{Marginal-Refs} for the $st^{2}$ model as well. We thus complete the 
analysis and extend the results obtained in \cite{Hotta}, \cite{GLS-1} and \cite{Cai-Pang}
for the particular case of $st^{2}$. 
\item  Final remarks, comments and future outlook are
given in the concluding Sect. \ref{Conclusion}.
\end{itemize}
\smallskip

\section{\label{st2}Basics of the $st^{2}$ Model}
Cubic special K\"ahler geometries in $N=2$, $d=4$ supergravities are a subset of the special K\"ahler geometries describing the $\sigma$-model of the scalar fields in the vector multiplets.
The distinguishing feature is the cubic prepotential function $F(X^\Lambda)$, which can arise in the large volume limit of Calabi-Yau compactifications of Type II superstrings or as reduction of minimal supergravity coupled to vector multiplets in $d=5$.

Using special coordinates $z^i=X^i/X^0=x^i-i \,y^i$ ($i=1,\ldots,n$), cubic special K\"ahler manifolds are described by a set of constants $d_{ijk}$, defining the holomorphic prepotential
\begin{eqnarray}
F(X) &=&\frac{1}{3!}d_{ijk}\frac{X^iX^jX^k}{X^0}=(X^0)^2f(z)\, ,
\label{prepotf0}\\[3mm]
f(z) &=&\frac{1}{3!}d_{ijk}z^i z^j z^k \, .
\end{eqnarray}

The $st^2$ model is a $\sigma$-model described by the coset manifold $\left[{\rm SU}(1,1)/{\rm U}(1)\right]^2$ with a cubic prepotential
\begin{equation}
	F(X) = \frac{X^1 (X^2)^2}{X^0},
\end{equation}
which falls in the general classification given in (\ref{prepotf0}) for $d_{122} = 2$.
The name of the model is a consequence of the expression of the prepotential in terms of the special coordinates:
\begin{equation}
	s = \frac{X^1}{X^0} \qquad \hbox{and} \qquad t = \frac{X^2}{X^0},
\end{equation}
which leads to $F(X)/(X^0)^2 = f(s,t) = s t^2$.

Here we start by recalling some of the basic facts of the $st^{2}$ model and hence fix our notations and conventions. The two complex moduli of the model can be  defined as
\begin{equation}
z^{1}\equiv x^{1}-iy^{1}\equiv s,~~z^{2}\equiv x^{2}-iy^{2}\equiv ~t
\label{boss-3-bis}
\end{equation}
with $x^{i},y^{i}\in \mathbb{R}_{0}^{+}$~\cite{Gilmore}. In \textit{special coordinates} (see e.g. \cite{4} and Refs. therein) the prepotential determining the relevant special K\"{a}hler
geometry reads
\begin{equation}
f=st^{2}.  \label{prepotential-stu}
\end{equation}
Working in special coordinates some of the geometric expressions take a simple and much elegant form, here for completeness we list the expressions for the K\"{a}hler potential, contravariant metric tensors, the non-vanishing component of the $C$-tensor, holomorphic central charge (also named superpotential) and BH effective potential ($i=1,2$ throughout) :
\begin{equation}
\begin{array}{l}
K=-ln\left[ -i(s-\overline{s})(t-\overline{t})^{2}\right]
\Rightarrow exp\left( -K\right) =8y^{1}(y^{2})^{2}; \\[0.5em]

g^{i\bar{j}}=-diag((s-\overline{s})^{2},\frac{1}{2}(t-\overline{t})^{2}%
; \\[0.5em]
C_{stt}=\frac{2i}{(s-\bar{s})(t-\bar{t})^{2}} ; \\[0.5em]
W(s,t)=q_{0}+q_{1}s+q_{2}t+p^{0}st^{2}-p^{1}t^{2}-2p^{2}st\,; \\[0.5em]
V_{BH}=\frac{i}{2(s-\bar{s})(t-\bar{t})^{2}} \cdot \left[
\,|W(s,t)|+|W(s,%
\bar{t})| + (p^{1}-p^{0}s)(t-\bar{t})^{2}\right] \\
\left[W(\bar{s},t)+W(\bar{s},\bar{t})+(p^{1}-p^{0}\bar{s})
(t-\bar{t})^{2}\right] + 2(|W(s,t)|^{2}+|W(s,\bar{t})|^{2}) .
\end{array}
\label{st2-geometry}
\end{equation}
Thus the covariantly holomorphic Central charge function for the $st^{2}$ model is
(see e.g. \cite{4} and Refs. therein)
\begin{eqnarray}
Z\left( s,t,\overline{s},\overline{t}\right) &\equiv
&e^{K/2}W\left( s,t\right) =  \notag \\
&=&\frac{1}{\sqrt{-i(s-\overline{s})(t-\overline{t})^{2}}}\left(
q_{0}+q_{1}s+q_{2}t+p^{0}st^{2}-p^{1}t^{2}-2p^{2}st\right) .
\notag \\
&&  \label{g-1}
\end{eqnarray}

The definition of the BH charges $p^{\Lambda }$ (magnetic) and $q_{\Lambda }$
(electric) ($\Lambda =0,1,2$ throughout), the effective $1$-dim.
(quasi-)geodesic Lagrangian of the $st^{2}$ model, and the corresponding Eqs.
of motion for the scalars can be computed following the same general method depicted in 
Subsects. 2.2 and 2.3, as well as in appendix A of \cite{GLS-1} (treating the case $D0-D4$ in detail), for the general $stu$ model and then making a degeneracy choice of $t=u$.
 
 According to the Bekenstein-Hawking entropy-area formula \cite{hawking2}, the
entropy of an extremal BH in the $st^{2}$ model in the Einsteinian
approximation can be written as follows:
\begin{equation}
S_{BH}=\frac{A_{H}}{4}=\pi \left. V_{BH}\right| _{\partial V_{BH}=0}=\pi
\sqrt{\left| \mathcal{I}_{4}\left( \Gamma \right) \right| },
\end{equation}
where the $\left( 2n_{V}+2\right) \times 1$ vector of BH charges
\begin{equation}
\Gamma \equiv \left( p^{\Lambda },q_{\Lambda }\right) ,  \label{Gamma}
\end{equation}
was introduced, $n_{V}$ denoting the number of Abelian vector multiplets
coupled to the supergravity one (in the case under consideration $n_{V}=2$).
Furthermore, $\mathcal{I}_{4}\left( \Gamma \right) $ denotes the unique
invariant of the $U$-duality group $G$, reading as
follows (see \textit{e.g.} Eq. (3.10) of \cite{BMOS-1}, and Refs. therein):
\begin{equation}
\mathcal{I}_{4}\left( \Gamma \right)=\mathcal{I}_{4},st^{2}\left( p,q \right)
=-(p^{0}q_{0}+p^{1}q_{1})^{2}+(2p^{1}p^{2}-p^{0}q_{2})(2p^{2}q_{0}+q_{1}q_{2})
 ,
\label{I4}
\end{equation}
More details regarding the relation of $\mathcal{I}_{4},st^{2}\left( p,q \right)$ and the so-called
Cayley's Hyperdeterminant \cite{duff1} can be found in Sect.5 of \cite{BMOS-1}.(see also Eqn. (3.6) of \cite{CDFY})
\smallskip

In the next three Sections we will discuss the explicit solutions of the
equations of motion of the scalars $s$ and $t$ in the dyonic background
of an extremal BH of the $st^{2}$ model, also named \textit{Attractor Flow
Equations}. We will consider only \textit{non-degenerate} attractor flows,
\textit{i.e.} those flows determining a regular Black Hole solution with non-vanishing 
area of the horizon in the Einsteinian approximation.
 
As mentioned above, $3$ classes of non-degenerate attractor flows exist in
the $st^{2}$ model: 

\begin{itemize}
\item  $\frac{1}{2}$-BPS (Sect. \ref{BPS-Flow});

\item  non-BPS $Z=0$ (Sect. \ref{Non-BPS-Z=0-Flow});

\item  non-BPS $Z\neq 0$ (Sects. \ref{Non-BPS-Z<>0-Flow}) and (\ref
{Analysis-Particular}).
\end{itemize}

\section{\label{BPS-Flow}The Most General $\frac{1}{2}$-BPS Attractor Flow}

The explicit expression of the attractor flow solution supported by the most
general $\frac{1}{2}$-BPS BH charge configuration in $\mathcal{N}=2$, $d=4$
\textit{ungauged} supergravity coupled to $n_{V}$ Abelian vector multiplets
(and exhibiting a unique $U$-invariant $\mathcal{I}_{4}$) is known after
\cite{Cvetic-Youm}-\nocite
{Tseytlin,Gauntlett,Bala,BPS-flow-1,BPS-flow-2,BPS-flow-3,Bates-Denef}\cite
{bala-foam} (as well as the third of Refs. \cite{Marginal-Refs}):
\begin{eqnarray}
exp\left[ -4U_{\frac{1}{2}-BPS}\left( \tau \right) \right] &=&\mathcal{I}%
_{4}\left( \mathcal{H}\left( \tau \right) \right) ;  \notag \\
z_{\frac{1}{2}-BPS}^{i}\left( \tau \right) &=&\frac{H^{i}\left( \tau \right)
+i\partial _{H_{i}}\mathcal{I}_{4}^{1/2}\left( \mathcal{H}\left( \tau
\right) \right) }{H^{0}\left( \tau \right) +i\partial _{H_{0}}\mathcal{I}%
_{4}^{1/2}\left( \mathcal{H}\left( \tau \right) \right) },  \label{BPS-sol}
\end{eqnarray}
where $\partial _{H_{i}}\equiv \frac{\partial }{\partial H_{i}}$, and the $%
\left( 2n_{V}+2\right) \times 1$($=6\times 1$ in the model under
consideration) symplectic vector
\begin{equation}
\mathcal{H}\left( \tau \right) \equiv \left( H^{\Lambda }\left( \tau \right)
,H_{\Lambda }\left( \tau \right) \right) ,  \label{H}
\end{equation}
was introduced, where $H^{\Lambda }\left( \tau \right) $ and $H_{\Lambda
}\left( \tau \right) $ are \textit{harmonic functions} defined as follows ($%
\tau \equiv \left( r_{H}-r\right) ^{-1}\in \mathbb{R}^{-}$):
\begin{equation}
\begin{array}{l}
H^{\Lambda }\left( \tau \right) \equiv p_{\infty }^{\Lambda }+p^{\Lambda
}\tau ; \\
\\
H_{\Lambda }\left( \tau \right) =q_{\Lambda ,\infty }+q_{\Lambda }\tau ,
\end{array}
\label{HH}
\end{equation}
such that $\mathcal{H}\left( \tau \right) $ can be formally rewritten as
\begin{equation}
\mathcal{H}\left( \tau \right) =\Gamma _{\infty }+\Gamma \tau .  \label{HHH}
\end{equation}
The asymptotical constants $\Gamma _{\infty }$ must satisfy the following
\textit{integrability conditions}:
\begin{equation}
\mathcal{I}_{4}\left( \Gamma _{\infty }\right) =1,~~\left\langle \Gamma
,\Gamma _{\infty }\right\rangle =0,  \label{constraints}
\end{equation}
where $\left\langle \cdot ,\cdot \right\rangle $ is the scalar product
defined by the $\left( 2n_{V}+2\right) \times \left( 2n_{V}+2\right) = (6\times6)$
symplectic metric. Under such conditions, the flow (\ref{BPS-sol}) is the
most general solution of the so-called $\frac{1}{2}$\textit{-BPS
stabilization Eqs. }(see \textit{e.g.} the recent treatment of \cite{K2-bis}%
):
\begin{equation}
\mathcal{H}^{T}\left( \tau \right) =2e^{K\left( z\left( \tau \right) ,%
\overline{z}\left( \tau \right) \right) }Im\left[ W\left( z\left( \tau
\right) ,\mathcal{H}\left( \tau \right) \right) \left(
\begin{array}{c}
\overline{X}^{\Lambda }\left( \overline{z}\left( \tau \right) \right) \\
\\
\overline{F}_{\Lambda }\left( \overline{z}\left( \tau \right) \right)
\end{array}
\right) \right] ,  \label{BPS-stabilization-Eqs.}
\end{equation}
obtained from the $\frac{1}{2}$\textit{-BPS Attractor Eqs.} (see \textit{e.g.%
} the treatment in \cite{AoB-book}, and Refs. therein)
\begin{equation}
\Gamma ^{T}=2e^{K\left( z,\overline{z}\right) }Im\left[ W\left( z,\Gamma
\right) \left(
\begin{array}{c}
\overline{X}^{\Lambda }\left( \overline{z}\right) \\
\\
\overline{F}_{\Lambda }\left( \overline{z}\right)
\end{array}
\right) \right]  \label{BPS-AEs}
\end{equation}
by simply replacing $\Gamma $ with $\mathcal{H}\left( \tau \right) $ (see
\textit{e.g.} \cite{BPS-flow-1} and Refs. therein). Consistently, Eq. (\ref
{BPS-AEs}) is the \textit{near-horizon} ($\tau \rightarrow -\infty $)
\textit{limit }of Eq. (\ref{BPS-stabilization-Eqs.}).

Moreover, the BH charge configurations supporting the $\frac{1}{2}$-BPS
attractors at the BH event horizon satisfy the following constraints,
defining the $\frac{1}{2}$-BPS orbit (see Appendix II of \cite{bellucci1} for a 
detailed discussion on this issue)
\begin{equation}
\mathcal{O}_{\frac{1}{2}-BPS}=\frac{\left( SU\left( 1,1\right) \right) ^{2}}{%
\left( U\left( 1\right) \right)}  \label{Wed-5}
\end{equation}
of the \textit{bi-fundamental} representation $\left( \mathbf{2},%
\mathbf{2}\right) $ of the $U$-duality group $\left( SU\left( 1,1\right)
\right) ^{2}$ \cite{bellucci1,BMOS-1}:
\begin{equation}
\begin{array}{l}
\mathcal{I}_{4}\left( \Gamma \right) >0; \\
\\
(p^{2})^{2}-p^{0}q_{1}\gtrless 0; \\
\\
2p^{1}p^{2}-p^{0}q_{2}\gtrless 0; \\
\end{array}
\label{Wed-1}
\end{equation}
Correspondingly, $\mathcal{H}\left( \tau \right) $ is constrained as follows
along the $\frac{1}{2}$-BPS attractor flow ($\forall \tau \in \mathbb{R}^{-}$%
):
\begin{equation}
\begin{array}{l}
\mathcal{I}_{4}\left( \mathcal{H}\left( \tau \right) \right) >0; \\
\\
(H^{2}\left( \tau \right))^{2}  - H^{0}\left( \tau \right)
H_{1}\left( \tau \right) \gtrless 0; \\
\\
2H^{1}\left( \tau \right) H^{2}\left( \tau \right) -H^{0}\left( \tau \right)
H_{2}\left( \tau \right) \gtrless 0.
\end{array}
\label{Wed-2}
\end{equation}

In the \textit{near-horizon limit} $\tau \rightarrow -\infty $, Eq. (\ref
{BPS-sol}) yields the \textit{purely charge-dependent}, \textit{critical}
expressions of the scalars at the BH event horizon.
In the same limit, the constraints (\ref{Wed-2})
consistently yield the constraints (\ref{Wed-1}).

Consistently with the analysis of \cite{Ceresole} performed for $stu$ model, the general $\frac{1}{2}$%
-BPS attractor flow solution (\ref{BPS-sol}) of the $st^{2}$ model can be
\textit{axion-free} only for the configurations $D0-D6,D0-D4$ (magnetic) and $D2-D6$ (electric).

As found in \cite{Ferrara-Gimon} and observed also in \cite{GLS-1}, an
immediate consequence of Eq. (\ref{BPS-sol}) is that $\Gamma _{\infty }$
satisfies the $\frac{1}{2}$\textit{-BPS Attractor Eqs. }\cite{BPS-flow-1}.
This determines a sort of \textit{``Attractor Mechanism at spatial
infinity'',} mapping the $4$ \textit{real} moduli $\left(
x^{1},x^{2},y^{1},y^{2}\right) $ into the $6$ \textit{real}
constants $\left( p_{\infty }^{1},p_{\infty }^{2}
,q_{1,\infty },q_{2,\infty }\right) $, arranged as $%
\Gamma _{\infty }$ and constrained by the $2$ \textit{real} conditions (\ref
{constraints}).

As noticed in \cite{GLS-1}, the absence of \textit{flat} directions in the $%
\frac{1}{2}$-BPS attractor flow (which is a general feature of $\mathcal{N}%
=2 $, $d=4$ \textit{ungauged} supergravity coupled to Abelian vector
multiplets, \textit{at least} as far as the metric of the scalar manifold is
strictly positive-definite $\forall \tau \in \mathbb{R}^{-}$ \cite{FGK}) is
crucial for the validity of the expression (\ref{BPS-sol}).\medskip

Now, by exploiting the \textit{first-order formalism} \cite{Fake-Refs} for $%
d=4$ extremal BHs \cite{Cer-Dal-1,ADOT-1} (see also \cite{FHM} and \cite
{Gnecchi-1}), one can compute the relevant BH parameters of the $\frac{1}{2}$%
-BPS attractor flow of the $st^{2}$ model starting from the expression of the $%
\frac{1}{2}$-BPS \textit{fake superpotential} $\mathcal{W}_{\frac{1}{2}-BPS}$%
. For instance, the \textit{ADM mass} and \textit{covariant scalar charges}
respectively read (see \textit{e.g.} the treatments in \cite{FHM} and \cite
{Gnecchi-1}):
\begin{eqnarray}
M_{ADM}\left( z_{\infty },\overline{z}_{\infty },\Gamma \right) &=&\mathcal{W%
}\left( z_{\infty },\overline{z}_{\infty },\Gamma \right) \equiv lim_{\tau
\rightarrow 0^{-}}\mathcal{W}\left( z\left( \tau \right) ,\overline{z}\left(
\tau \right) ,\Gamma \right) ;  \label{d-1} \\
&&  \notag \\
\Sigma _{i}\left( z_{\infty },\overline{z}_{\infty },\Gamma \right)
&=&\left( \partial _{i}\mathcal{W}\right) \left( z_{\infty },\overline{z}%
_{\infty },\Gamma \right) \equiv lim_{\tau \rightarrow 0^{-}}\left( \partial
_{i}\mathcal{W}\right) \left( z\left( \tau \right) ,\overline{z}\left( \tau
\right) ,\Gamma \right) ,  \label{d-2}
\end{eqnarray}
where the subscript \textit{``}$\infty $'' denotes the evaluation at the
moduli at spatial infinity ($r\rightarrow \infty \Leftrightarrow \tau
\rightarrow 0^{-}$). Notice that Eq. (\ref{d-1}) provides, within the
considered \textit{first-order formalism}, an alternative (eventually
simpler) formula for the computation of $M_{ADM}$, with respect to the
general definition in terms of the \textit{warp factor} $U$ (see \textit{e.g.%
} \cite{FGK}):
\begin{equation}
M_{ADM}=lim_{\tau \rightarrow 0^{-}}\frac{dU\left( \tau \right) }{d\tau }.
\end{equation}
Recalling that for all $\mathcal{N}=2$, $d=4$ \textit{ungauged}
supergravities it holds that $\mathcal{W}_{\frac{1}{2}-BPS}=\left| Z\right| $%
, Eqs. (\ref{g-1}) and (\ref{d-1}) yield the following expressions of the
\textit{ADM mass} of the $\frac{1}{2}$-BPS attractor flow of the $st^{2}$
model:
\begin{eqnarray}
M_{ADM,\frac{1}{2}-BPS}\left( z_{\infty },\overline{z}_{\infty },\Gamma
\right) &\equiv &lim_{\tau \rightarrow 0^{-}}\left| Z\right| \left( z\left(
\tau \right) ,\overline{z}\left( \tau \right) ,\Gamma \right) =  \notag \\
&=&\frac{\left| q_{0}+q_{1}s_{\infty }+q_{2}t_{\infty
}+p^{0}s_{\infty }t_{\infty }^{2}-p^{1}t_{\infty }^{2}
-2p^{2}s_{\infty }t_{\infty }\right| }{\sqrt{%
-i(s_{\infty }-\overline{s}_{\infty })(t_{\infty }-\overline{t}_{\infty
})^{2}}}\,.  \notag \\
&&  \label{M-BPS-gen}
\end{eqnarray}
Thus we have,
 $$\mathcal{W}_{\frac{1}{2}-BPS}=\left| Z\right| $$
 Equation (\ref{M-BPS-gen}) yields that the \textit{marginal bound} \cite
{Marginal-Refs} is not \textit{saturated} by $\frac{1}{2}$-BPS states,
because $M_{ADM,\frac{1}{2}-BPS}$ is \textit{not} equal to the sum of the
\textit{ADM masses} of the $D$-branes with appropriate fluxes (for further
detail, see the discussion in \cite{GLS-1}).

Concerning the \textit{(covariant) scalar charges} of the $\frac{1}{2}$-BPS
attractor flow of the $st^{2}$ model, they can be straightforwardly computed by
using Eqs. (\ref{g-1}) and (\ref{d-2}):
\begin{gather}  \label{Sigma-BPS-s}
\Sigma _{s,\frac{1}{2}-BPS}\left( z_{\infty },\overline{z}_{\infty },\Gamma
\right) \equiv lim_{\tau \rightarrow 0^{-}}\left( \partial _{s}\left|
Z\right| \right) \left( z\left( \tau \right) ,\overline{z}\left( \tau
\right) ,\Gamma \right) \nonumber\\
= lim_{\tau \rightarrow 0^{-}}\frac{\left[ \left(
\partial _{s}Z\right) \overline{Z}+Z\partial _{s}\overline{Z}\right] }{%
2\left| Z\right| }\left( z\left( \tau \right) ,\overline{z}\left( \tau
\right) ,\Gamma \right) \notag \\
= lim_{\tau \rightarrow 0^{-}}\frac{e^{K/2}}{2}\left[\left( \partial
_{s}K\right) \left| W\right| +\left( \partial _{s}W\right) \sqrt{\frac{%
\overline{W}}{W}}\right] \left( z\left( \tau \right) ,\overline{z}\left(
\tau \right) ,\Gamma \right)  \notag \\
=\frac{1}{2\sqrt{-i(s_{\infty }-\overline{s}_{\infty })(t_{\infty }-%
\overline{t}_{\infty })^{2}}}\cdot  \notag \\
\cdot \left[ \frac{\left| q_{0}+q_{1}s_{\infty }+q_{2}t_{\infty
}+p^{0}s_{\infty }t_{\infty }^{2}-p^{1}t_{\infty
}^{2}-2p^{2}s_{\infty }t_{\infty } \right|
}{-(s_{\infty }-\overline{s}_{\infty })}\right. +  \notag \\
\notag \\
+\left( q_{1}+p^{0}t_{\infty }^{2}-2p^{2}t_{\infty }\right) \cdot  \notag \\
\left. \cdot \sqrt{\frac{q_{0}+q_{1}\overline{s}_{\infty }+q_{2}\overline{t}%
_{\infty }+p^{0}\overline{s}_{\infty }\overline{t%
}_{\infty }^{2}-p^{1}\overline{t}_{\infty }^{2}%
-2p^{2}\overline{s}_{\infty }\overline{t}_{\infty }%
\,}{q_{0}+q_{1}s_{\infty }+q_{2}t_{\infty }+ p^{0}s_{\infty
}t_{\infty }^{2}-p^{1}t_{\infty }^{2}-2p^{2}s_{\infty }t_{\infty
}}}\right]
.  \notag \\
\end{gather}

\begin{gather}  \label{Sigma-BPS-t}
\Sigma _{t,\frac{1}{2}-BPS}\left( z_{\infty },\overline{z}_{\infty },\Gamma
\right) \equiv lim_{\tau \rightarrow 0^{-}}\left( \partial _{t}\left|
Z\right| \right) \left( z\left( \tau \right) ,\overline{z}\left( \tau
\right) ,\Gamma \right) \nonumber\\
= lim_{\tau \rightarrow 0^{-}}\frac{\left[ \left(
\partial _{t}Z\right) \overline{Z}+Z\partial _{t}\overline{Z}\right] }{%
2\left| Z\right| }\left( z\left( \tau \right) ,\overline{z}\left( \tau
\right) ,\Gamma \right)  \notag \\
=lim_{\tau \rightarrow 0^{-}}\frac{e^{K/2}}{2}\left[\left( \partial
_{t}K\right) \left| W\right| +\left( \partial _{t}W\right) \sqrt{\frac{%
\overline{W}}{W}}\right] \left( z\left( \tau \right) ,\overline{z}\left(
\tau \right) ,\Gamma \right) \notag \\
=\frac{1}{2\sqrt{-i(s_{\infty }-\overline{s}_{\infty })(t_{\infty }-%
\overline{t}_{\infty })^{2}}}\cdot  \notag \\
\cdot \left[ \frac{\left| q_{0}+q_{1}s_{\infty }+q_{2}t_{\infty
}+p^{0}s_{\infty }t_{\infty }^{2}-p^{1}t_{\infty
}^{2}-2p^{2}s_{\infty }t_{\infty }\right|
}{-\frac{1}{2}(t_{\infty }-\overline{t}_{\infty })}\right. +  \notag \\
\notag \\
+\left( q_{2}+2p^{0}s_{\infty }t_{\infty }-2p^{1}t_{\infty
}-2p^{2}s\,_{\infty
}\right) \cdot  \notag \\
\left. \cdot \sqrt{\frac{q_{0}+q_{1}\overline{s}_{\infty }+q_{2}\overline{t}%
_{\infty }+p^{0}\overline{s}_{\infty }\overline{t%
}_{\infty }^{2}-p^{1}\overline{t}_{\infty }^{2}%
-2p^{2}\overline{s}_{\infty }\overline{t}_{\infty }%
\,}{q_{0}+q_{1}s_{\infty }+q_{2}t_{\infty }+p^{0}s_{\infty
}t_{\infty }^{2}-p^{1}t_{\infty }^{2}-2p^{2}s_{\infty }t_{\infty
}}}\right]
.  \notag \\
\end{gather}\smallskip

\section{\label{Non-BPS-Z=0-Flow}The Most General Non-BPS $Z=0$ Attractor
Flow} 

Let us now investigate the non-BPS $Z=0$ case.

As shortly noticed in \cite{GLS-1}, in spite of the fact that this attractor
flow is non-supersymmetric, it has many common features with the
supersymmetric ($\frac{1}{2}$-BPS) case.

As yielded by the analysis of \cite{BMOS-1}, the non-BPS $Z=0$ horizon
attractor solutions can be obtained from $\frac{1}{2}$-BPS ones simply by
changing the signs of  the imaginary parts of the second moduli (\textit{dilatons%
}) and consistently imposing specific constraints on BH charges. 
Hence one has the only possible choice to flip the dilatons as follows:
\begin{equation}
y^{1}\rightarrow y^{1},\;y^{2}\rightarrow -y^{2}.
\label{Wed-6}
\end{equation}
This yields the following constraints on the BH charge configurations
supporting the non-BPS $Z=0$ attractors at the BH event horizon ($\tau
\rightarrow -\infty $) \cite{BMOS-1}:
\begin{equation}
\begin{array}{l}
\mathcal{I}_{4}\left( \Gamma \right) >0; \\
\\
(p^{2})^{2}-p^{0}q_{1}\lessgtr 0; \\
\\
2p^{1}p^{2}-p^{0}q_{2}\gtrless 0.
\end{array}
\label{Wed-3}
\end{equation}
The constraints (\ref{Wed-3}) defines the non-BPS $Z=0$ orbit of the \textit{%
bi-fundamental} representation $\left( \mathbf{2},\mathbf{2}%
\right) $ of the $U$-duality group $\left( SU\left( 1,1\right) \right) ^{2}$
(see Appendix II of \cite{bellucci1})
\begin{equation}
\mathcal{O}_{non-BPS,Z=0}=\frac{\left( SU\left( 1,1\right) \right) ^{2}}{%
\left( U\left( 1\right) \right)}.  \label{Wed-4}
\end{equation}
Notice that such an orbit shares the same coset expression of $\mathcal{O}_{%
\frac{1}{2}-BPS}$ given by Eq. (\ref{Wed-5}). However, they do \textit{not}
coincide, but instead they are two \textit{separated} branches of a \textit{%
disconnected} manifold, classified by the local value of the function $%
sgn\left( \left| Z\right| ^{2}-\left| D_{s}Z\right| ^{2}\right)$

The same holds \textit{all along the attractor flow}, \textit{i.e.} $\forall
\tau \in \mathbb{R}^{-}$. Indeed, the most general non-BPS $Z=0$ attractor
flow can be obtained by taking the most general $\frac{1}{2}$-BPS attractor
flow, and flipping any one out of the two dilatons. Thus, by taking Eq. (%
\ref{BPS-sol}) and flipping the dilatons as given by Eq. (\ref{Wed-6}), one
achieves the following result:

\begin{eqnarray}
exp\left[ -4U_{non-BPS,Z=0}\left( \tau \right) \right] &=&\mathcal{I}%
_{4}\left( \mathcal{H}\left( \tau \right) \right) ;  \notag \\
&&  \notag \\
z_{non-BPS,Z=0}^{1}&=&\frac{H^{\Lambda }(\tau )H_{\Lambda
}(\tau
)-2H^{1}(\tau )H_{1}(\tau )-i\mathcal{I}_{4}^{1/2}(\mathcal{H}(\tau ))}{2%
\left[ (H^{2}(\tau ))^{2}-H^{0}(\tau )H_{1}(\tau )\right] }=z_{\frac{1%
}{2}-BPS}^{1}(\tau );  \notag \\
&&  \notag \\
z_{non-BPS,Z=0}^{2}&=&\frac{H^{\Lambda }(\tau )H_{\Lambda
}(\tau
)-2H^{2}(\tau )H_{2}(\tau )+i\mathcal{I}_{4}^{1/2}(\mathcal{H}(\tau ))}{2%
\left[ H^{1}(\tau )H^{2}(\tau )-H^{0}(\tau )H_{2}(\tau )\right] }=\overline{%
z_{\frac{1}{2}-BPS}^{2}}(\tau );  .  \label{non-BPS-Z=0-flow}
\end{eqnarray}
This is the most general expression of the non-BPS $Z=0$ attractor flow, in
the \textit{``polarization'' }given by Eq. (\ref{Wed-6}).
Consistently with the constraints (\ref{Wed-3}), $\mathcal{H}\left( \tau
\right) $ is constrained as follows along the non-BPS $Z=0$ attractor flow ($%
\forall \tau \in \mathbb{R}^{-}$):
\begin{equation}
\begin{array}{l}
\mathcal{I}_{4}\left( \mathcal{H}\left( \tau \right) \right) >0; \\
\\
(H^{2}\left( \tau \right))^{2}  -H^{0}\left( \tau \right)
H_{1}\left( \tau \right) \lessgtr 0; \\
\\
2H^{1}\left( \tau \right) H^{2}\left( \tau \right) -H^{0}\left( \tau
\right) H_{2}\left( \tau \right) \gtrless 0 .
\end{array}
\label{Wed-7}
\end{equation}
In the \textit{near-horizon limit} $\tau \rightarrow -\infty $, Eq. (\ref
{non-BPS-Z=0-flow}) yields the \textit{purely charge-dependent}, \textit{%
critical} expressions of the scalars at the BH event horizon, given by Eq.
(3.9) of \cite{BMOS-1}. In the same limit, the constraints (\ref{Wed-7})
consistently yield the contraints (\ref{Wed-3}). The \textit{integrability
conditions} (\ref{constraints}) clearly hold also in this case.

Consistently with the analysis of \cite{Ceresole}, the general non-BPS $Z=0$
attractor flow solution (\ref{non-BPS-Z=0-flow}) of the $st^{2}$ model can be
\textit{axion-free} only for the configurations $D0-D6,D0-D4$ (magnetic) and
$D2-D6$ (electric).

A consequence of Eq. (\ref{non-BPS-Z=0-flow}) is that $\Gamma _{\infty }$
satisfies the \textit{non-BPS }$\mathit{Z=0}$ \textit{Attractor Eqs. }(see
\textit{e.g.} \cite{AoB-book} ). Analogously to what happens
for the $\frac{1}{2}$-BPS attractor flow, this determines a sort of \textit{%
``Attractor Mechanism at spatial infinity''}.

Analogously to what happens in the $\frac{1}{2}$-BPS case, the absence of
\textit{flat} directions in the non-BPS $Z=0$ attractor flow  for the
$st^{2}$ model is crucial for the validity of the expression (\ref{non-BPS-Z=0-flow}).

By exploiting the strict relation with the $\frac{1}{2}$-BPS attractor flow,
one can also determine the explicit expression of the \textit{fake
superpotential }$\mathcal{W}_{non-BPS,Z=0}$ for the non-BPS $Z=0$ attractor
flow. Considering the absolute value of the $\mathcal{N}=2$, $d=4$ \textit{%
central charge function} $Z$ given by Eq. (\ref{g-1}) and flipping one
out of the two dilatons in the \textit{``polarization'' }given by Eq. (\ref
{Wed-6}), one obtains the following non-BPS $Z=0$ \textit{fake superpotential%
} (notice that $K$, as given by the first Eq. of (\ref{st2-geometry}), is
invariant under such a flipping):
\begin{eqnarray}
\mathcal{W}_{non-BPS,Z=0,s} &=&e^{K/2}\left| q_{0}+q_{1}s+q_{2}\overline{t}%
+p^{0}s\overline{t}^{2}-p^{1}\overline{t}^{2}%
-2p^{2}s\overline{t}\right| =  \notag \\
&=&\left| Z\left( s,\overline{t}\right) \right| =\mathcal{W}_{%
\frac{1}{2}-BPS}\left( s,\overline{t}\right) ,  \label{Ws}
\end{eqnarray}
where the subscript \textit{``}$s$\textit{'' }denotes the modulus untouched
by the considered flipping of dilatons; in the last step we used the
aforementioned fact that for all $\mathcal{N}=2$, $d=4$ \textit{ungauged}
supergravities it holds that $\mathcal{W}_{\frac{1}{2}-BPS}=\left| Z\right| $

Like the triality symmetry in the $stu$ model here in the case of $st^{2}$ model
there is no equivalent flipping of the moduli like
\begin{equation}
y^{1}\rightarrow -y^{1},\;y^{2}\rightarrow y^{2},
\label{Wed-7}
\end{equation}
as the triality symmetry is completely broken once one chooses the last two moduli
to be equal in case of $stu$ model to generate the $st^{2}$ model, and thus it is 
possible to have a new symmetry for the $st^{2}$ model like this: 
\begin{equation}
y^{1}\rightarrow -y^{1},\;y^{2}\rightarrow e^{i\frac{\pi}{2}}y^{2}.
\label{new}
\end{equation}
such that under this symmetry transformation the K\"{a}hler potential (given by the first of Eqn. \ref{st2-geometry}) remains invariant.

Now, as shown in \cite{stuunveiled}, by exploiting the \textit{first-order formalism} 
\cite{Fake-Refs} for $%
d=4$ extremal BHs \cite{Cer-Dal-1,ADOT-1} (see also \cite{FHM} and \cite
{Gnecchi-1}), one can compute the relevant BH parameters of the non-BPS $Z=0$
attractor flow of the $stu$ model starting from the expression of the
non-BPS $Z=0$ \textit{fake superpotential} $\mathcal{W}_{non-BPS,Z=0}$ . 
The choice of \textit{``}$s$%
\textit{-polarization''}, \textit{``}$t$\textit{-polarization''} or \textit{%
``}$u$\textit{-polarization'' } was immaterial, due to the underlying \textit{%
triality symmetry} of the moduli $s$, $t$ and $u$. Thus, without loss of
generality, they choose to perform computations in the \textit{``}$s$\textit{%
-polarization''} (equivalent results in the other two \textit{%
``polarizations''} can be obtained by cyclic permutations of the moduli).
But in our case of $st^{2}$ model because of lack of triality symmetry, one can't use the cyclic permutation of the moduli to derive results for all of them just by computing it for one. What one needs is to compute each of them separately.

The ADM mass of the non-BPS $Z=0$ attractor flow of the $st^{2}$
model is :
\begin{eqnarray}
M_{ADM,non-BPS,Z=0}\left( z_{\infty },\overline{z}_{\infty },\Gamma \right)
&\equiv &lim_{\tau \rightarrow 0^{-}}\mathcal{W}_{non-BPS,Z=0,s}\left(
z\left( \tau \right) ,\overline{z}\left( \tau \right) ,\Gamma \right) =
\notag \\
&=&lim_{\tau \rightarrow 0^{-}}\left| Z\left( s\left( \tau \right) ,%
\overline{t}\left( \tau \right)  \right)
\right| =  \notag \\
&=&\frac{\left| q_{0}+q_{1}s_{\infty }+q_{2}\overline{t}_{\infty }+%
+p^{0}s_{\infty }\overline{t}_{\infty }^{2}%
p^{1}\overline{t}_{\infty }^{2}-2p^{2}s_{\infty }%
\overline{t}_{\infty }\right| }{\sqrt{%
-i(s_{\infty }-\overline{s}_{\infty })(t_{\infty }-\overline{t}_{\infty
})^{2}}}\,.  \notag \\
&&  \label{M-non-BPS-Z=0-gen}
\end{eqnarray}

Eq. (\ref{M-non-BPS-Z=0-gen}) yields that the \textit{marginal bound} \cite
{Marginal-Refs} is not \textit{saturated} by non-BPS $Z=0$ states, because $%
M_{ADM,non-BPS,Z=0}$ is \textit{not} equal to the sum of the \textit{ADM
masses} of the $D$-branes with appropriate fluxes (for further detail, see
the discussion in \cite{GLS-1}). This is actually expected, due to the
strict similarity, discussed above, between $\frac{1}{2}$-BPS and non-BPS $%
Z=0$ attractor flows in the considered $st^{2}$ model; such a similarity can be
explained by noticing that both of the flows can be uplifted to the \textit{%
same} $\frac{1}{8}$-BPS \textit{non-degenerate} attractor flow of $\mathcal{N%
}=8$, $d=4$ supergravity (see \textit{e.g.} the discussion in \cite{BMOS-1}).

Concerning the covariant scalar charges of the non-BPS $Z=0$ attractor flow of
the $st^{2}$ model  they can be straightforwardly computed
(in the \textit{``}$s$\textit{-polarization''}, and (in the \textit{``}$t$\textit{-polarization''},
separately by using Eqs. (\ref{Ws}) and (\ref{d-2}), but here we write the expression for the scalar charge taking into account only the \textit{``}$s$\textit{-polarization''} as:
\begin{gather}
\Sigma _{s,non-BPS,Z=0}\left( z_{\infty },\overline{z}_{\infty },\Gamma
\right) \equiv lim_{\tau \rightarrow 0^{-}}\left( \partial _{s}\mathcal{W}%
_{non-BPS,Z=0,s}\right) \left( z\left( \tau \right) ,\overline{z}\left( \tau
\right) ,\Gamma \right) =  \notag \\
=lim_{\tau \rightarrow 0^{-}}\partial _{s}\left| Z\left( s\left(
\tau \right) ,\overline{t}\left( \tau \right)
\right) \right| =  \notag \\
=lim_{\tau \rightarrow 0^{-}}\frac{e^{K/2}}{2}\left[\left( \partial
_{s}K\right) \left| W\left( s,\overline{t}\right) \right| +\left(
\partial _{s}W\left( s,\overline{t}\right) \right)
\sqrt{\frac{\overline{W}\left( \overline{s},t\right) }{W\left( s,\overline{%
t}\right) }}\right] =  \notag \\
=\frac{1}{2\sqrt{-i(s_{\infty }-\overline{s}_{\infty })(t_{\infty }-%
\overline{t}_{\infty })^{2}}}\cdot  \notag \\
\cdot \left[ \frac{\left| q_{0}+q_{1}s_{\infty
}+q_{2}\overline{t}_{\infty
}+p^{0}s_{\infty }\overline{t}_{\infty }^{2}%
-p^{1}\overline{t}_{\infty }^{2}-2p^{2}s_{\infty }\overline{t}_{\infty }%
\right| }{-(s_{\infty }-\overline{s}_{\infty })}\right. +  \notag \\
\notag \\
+\left( q_{1}+p^{0}\overline{t}_{\infty }^{2}-2p^{2}%
\overline{t}_{\infty }\right) \cdot  \notag \\
\left. \cdot \sqrt{\frac{q_{0}+q_{1}\overline{s}_{\infty
}+q_{2}t_{\infty }+p^{0}\overline{s}_{\infty }t_{\infty
}^{2}-p^{1}t_{\infty }^{2}-2p^{2}\overline{s}_{\infty }t_{\infty }%
\,}{q_{0}+q_{1}s_{\infty }+q_{2}\overline{t}%
_{\infty }+p^{0}s_{\infty }\overline{t}_{\infty }^{2}%
-p^{1}\overline{t}_{\infty }^{2}-2p^{2}s_{\infty }\overline{t}_{\infty }%
}}\right] \notag \\
. \label{Sigma-non-BPS-Z=0-s}
\end{gather}

\section{\label{Non-BPS-Z<>0-Flow}The most General Non-BPS $Z\neq 0$
Attractor Flow}
All the features holding for $\frac{1}{2}$-BPS and non-BPS $Z=0$ attractor
flows (respectively treated in Sects. \ref{BPS-Flow} and \ref
{Non-BPS-Z=0-Flow}) do not directly hold for the non-BPS $Z\neq 0$ attractor
flow, which actually turns out to be rather different from (and structurally
much more intricate than) the other two attractor flows.

As mentioned in the Introduction, the non-BPS $Z\neq 0$ attractor flow of
the $stu$ model has been already considered in literature in particular
cases, namely for the $D0-D4$ (\textit{magnetic}) \cite{Hotta,GLS-1}, $D0-D6$
\cite{GLS-1}, $D2-D6$ (\textit{electric}) \cite{K2-bis,Cai-Pang} $D0-D2-D4$ (%
\textit{magnetic }with $D2$) \cite{Cai-Pang}, $D0-D2-D4-D6$ (without $B$%
\textit{-fields}) \cite{K2-bis} supporting BH charge
configurations.\smallskip

In the present Section we determine the explicit expression of the non-BPS $%
Z\neq 0$ attractor flow for \textit{the most general} supporting BH charge
configuration, with \textit{all} electric and magnetic charges switched on,
namely for the non-BPS $Z\neq 0$-supporting branch of the $D0-D2-D4-D6$
configuration. Thence, as already done for $\frac{1}{2}$-BPS and non-BPS $%
Z=0 $ attractor flows, by exploiting the \textit{first order (fake
supergravity) formalism} \cite{Fake-Refs, Cer-Dal-1,ADOT-1}, we compute the
\textit{ADM masses} as well as the \textit{covariant scalar charges}, and
study the issue of \textit{marginal stability }\cite{Marginal-Refs},
completing and refining the treatment given in \cite{Hotta,GLS-1,Cai-Pang} but for $st^{2}$
 model as an illustrative case.

\subsection{\label{D0-D6}The $ D0-D6$ solution with B-fields: }
\subsubsection{\label{U-Duality-Transform}$U$-Duality Transformations
along the Orbit $\mathcal{O}_{non-BPS,Z\neq 0}$}

In order to derive the explicit expression of the non-BPS $Z\neq 0$
attractor flow when all BH charges are non-vanishing, we exploit a method
already used in \cite{K2-bis}, \cite{GLS-1} and \cite{Cai-Pang}, based on
performing suitable symplectic transformations along the relevant (\textit{%
i.e.} supporting) charge orbit of the $U$-duality group. In Eqs. (\ref{Wed-5}%
) and (\ref{Wed-4}) we recalled the form of the $\frac{1}{2}$-BPS- and
non-BPS $Z=0$- supporting BH charge orbits of the \textit{bi-fundamental}
representation $\left( \mathbf{2},\mathbf{2}\right) $ of the $U$%
-duality group $G$ (given by Eq.(\ref{GG})) of the $st^{2}$ model. 
The corresponding non-BPS $%
Z\neq 0$-supporting BH charge orbit reads \cite{bellucci1}
\begin{equation}
\mathcal{O}_{non-BPS,Z\neq 0}=\frac{\left( SU\left( 1,1\right) \right) ^{2}}{%
\left( SO\left( 1,1\right) \right)},  \label{a-3}
\end{equation}
defined by the constraint
\begin{equation}
\mathcal{I}_{4}\left( \Gamma \right) <0.  \label{a-2}
\end{equation}

As done in \cite{GLS-1} and \cite{Cai-Pang}, for the $stu$ model, in order to perform a
symplectic transformation along the charge orbit $\mathcal{O}_{non-BPS,Z\neq
0}$ of the $\left( \mathbf{2},\mathbf{2}\right) $ representation
of the $U$-duality, we exploit the complete factorization of the
special K\"{a}hler manifold $\left( \frac{SU\left( 1,1\right) }{U\left(
1\right) }\right) ^{2}$,
which allows one to deal with the product of two distinct $2\times 2$
matrices of $SL\left( 2,\mathbb{R}\right) $, rather than with a unique matrix of the $U$-duality group embedded in the relevant
symplectic group $Sp\left( 6,\mathbb{R}\right) $.

The first step is to perform an $Sp\left( 6,\mathbb{R}\right) $%
-transformation from the basis $\left( p^{\Lambda },q_{\Lambda }\right) $ to
a basis $\mathcal{A}_{ab}$ ($a,b=0,1$ throughout) of BH charges expicitly
transforming under the $\left( \mathbf{2},\mathbf{2}\right) $ of
the $U$-duality. Such a transformation is similar to Eq. (5.1) of \cite{GLS-1} 
applied for the $stu$ case.
(equivalent to Eq. (3.5) of the second Ref. of \cite{duff1}; see also
Section 5 of \cite{BMOS-1}). The explicit action of a generic symplectic 
transformation of the $U$-duality on the BH charges $\mathcal{A}_{ab}$ is
given by,

\begin{eqnarray}
\mathcal{A}_{a^{\prime }b^{\prime }}^{\prime } &=&\left(
M_{1}\right) _{a^{\prime }}^{~a}\left( M_{2}\right) _{b^{\prime
}}^{~b}a_{ab};  \label{boss-1} \\
&&  \notag \\
M_{i} &\equiv &\left(
\begin{array}{cc}
a_{i} & b_{i} \\
c_{i} & d_{i}
\end{array}
\right)\in SL\left( 2,\mathbb{R}\right) ,~det\left( M_{i}\right) =1,~\forall i=1,2  \label{boss-2}
\end{eqnarray}
where each matrix pertains to the degrees of freedom of only one modulus (%
\textit{e.g.} $M_{1}$ to $s$, $M_{2}$ to $t$). The
transformation  (\ref{boss-1})-(\ref{boss-2}) of $\left( SL\left( 2,\mathbb{R}%
\right) \right) ^{2}\subset Sp\left( 6,\mathbb{R}\right) $ induces also a
\textit{fractional linear} (\textit{M\"{o}bius}) transformation on the
moduli $z^{i}$ as follows (no summation on repeated indices; also recall Eq. (\ref
{boss-3-bis})):

\begin{equation}
z^{\prime i}=\frac{a_{i}z^{i}+b_{i}}{c_{i}z^{i}+d%
_{i}}.  \label{boss-3}
\end{equation}

As done in \cite{GLS-1} and \cite{Cai-Pang} for the $stu$ model, we use the configuration $D0-D6$ as \textit{``pivot''} in order to perform the transformation (\ref{boss-1})-(%
\ref{boss-3}). Indeed, such a BH charge configuration supports only non-BPS $%
Z\neq 0$ attractors, as it can be easily realized by computing the
corresponding \textit{quartic} $U$-invariant, given by Eq. (\ref{I4}): $%
\mathcal{I}_{4}\left( \Gamma _{D0-D6}\right) <0$ . 
Thus, we want to transform from the configuration $D0-D6$
(corresponding to charges $\left( q_{0},p^{0}\right) $, which we denote here
$\left( q,p\right) $) to the most general configuration $D0-D2-D4-D6$,
corresponding to all BH charges switched on: $\left(
q_{0},q_{i},p^{i},p^{0}\right) $. By exploiting the transformation (\ref
{boss-1})-(\ref{boss-3}), the parameters $\frak{a}_{i},\frak{b}_{i},\frak{c}%
_{i},\frak{d}_{i}$ of the $M_{i}$s dualizing from $D0-D6$ to $D0-D2-D4-D6$
must satisfy the following set of constraints:

\begin{eqnarray}
\label{constraintsD0-D6}
-q_{0} &=&-a_{1}(a_{2})^{2}q+b_{1}(b_{2})^{2}%
p;  \notag \\
&&  \notag \\
0 &=&-c_{1}(a_{2})^{2}q+%
d_{1}(b_{2})^{2}p;  \notag \\
&&  \notag \\
0 &=&-c_{2}a_{1}a_{2}q+%
b_{2}d_{1}d_{2}p;  \notag \\
&&  \notag \\
p^{1} &=&-a_{1}(c_{2})^{2}q+%
b_{1}(d_{2})^{2}p;  \notag \\
&&  \notag \\
p^{2} &=&-a_{2}c_{1}c_{2}q+b_{2}d_{1}d_{2}%
p;  \notag \\
&&  \notag \\
0 &=&-c_{1}(c_{2})^{2}q+d_{1}(d_{2})^{2}%
p; \label{a-1}
\end{eqnarray}
Notice that the system (\ref{constraintsD0-D6}) admits solutions iff\ the condition 
(\ref{a-2}) is met; this implies the transformations (\ref{boss-1})-(\ref{boss-3}) to belong 
to the $U$%
-duality orbit $\mathcal{O}_{non-BPS,Z\neq 0}$ given by Eq. (\ref{a-3}). The
sign of the BH charges $q$ and $p$ is actually irrelevant for the condition (%
\ref{a-2}) to be satisfied; thus, without loss of any generality, one can
choose \textit{e.g.} $q>0$, $p>0$. Within such a choice, the explicit form
of the matrices $M_{i}$s under consideration (and of their inverse) reads as
follows:
\begin{eqnarray}
M_{i} &=&-\frac{1 }{\sqrt{ 2\lambda_{0}\varrho _{i} }}\left(
\begin{array}{cc}
\varrho _{i}\lambda_{0} & -\varrho _{i} \\
\lambda_{0} & 1
\end{array}
\right) \Leftrightarrow M_{i}^{-1}=-\frac{1 }{\sqrt{ 2\lambda_{0}\varrho _{i} }}%
\left(
\begin{array}{cc}
1 & \varrho _{i} \\
-\lambda_{0} & \varrho_{i}\lambda_{0}
\end{array}
\right) ;  \label{b-1'} 
\end{eqnarray}
 
Now we define all parameters of the matrices $M_{i}$ and their inverse: 
\begin{eqnarray}
\lambda &\equiv &\lambda_{0}\left[ \frac{%
2p^{1}(p^{2})^{2}+p^{0}\left( \sqrt{|\mathcal{I}_{4}|}-p^{\Lambda }q_{\Lambda
}\right) }{2p^{1}(p^{2})^{2}-p^{0}\left( \sqrt{|\mathcal{I}_{4}|}-p^{\Lambda
}q_{\Lambda }\right) }\right] ^{1/3} \in \mathbb{R};  \;\;\;\ \lambda_{0} = \left(\frac{p}{q}\right)^{\frac{1}{3}}\label{b-2'} \\
&&  \notag \\
\varrho _{i} &\equiv &\frac{\sqrt{|\mathcal{I}_{4}|}-p^{\Lambda }q_{\Lambda
}+2p^{i}q_{i}}{\epsilon _{ijk}p^{j}p^{k}-2p^{0}q_{i}}\in \mathbb{R}~\text{(no sum on
}i\text{)}.  \label{b-3'} 
\end{eqnarray}

For the $st^{2}$ model the unique quartic invariant reads as follows:
\begin{equation}
\mathcal{I}_{4},st^{2}\left( p,q \right)
=-(p^{0}q_{0}+p^{1}q_{1})^{2}+(2p^{1}p^{2}-p^{0}q_{2})(2p^{2}q_{0}+q_{1}q_{2})
\end{equation}

Thus the two values of $\varrho$  for the $D0-D6$ configuration with B fields are given by: 
\begin{eqnarray}
\varrho _{1} &\equiv &2\sqrt{\frac{q_{0}p^{1}}{(p^{2})^{2}}};
\label{b-3}\\
&&  \notag \\
&&  \notag \\
\varrho _{2} &\equiv &2\sqrt{\frac{q_{0}}{p^{1}}}; \label{b-4}
\end{eqnarray}

\subsubsection{\label{D0-D6}$D0-D6$: the Most General Flow and
\textit{Fake Superpotential}}
The most general non-BPS $Z\neq 0$ attractor flow in the $D0-D6$
configuration reads as follows:
\begin{eqnarray}
exp\left[ -4U_{non-BPS,Z\neq 0}\left( \tau \right) \right] &=&\left[
a-\left( -\mathcal{I}_{4}\right) ^{1/4}\tau \right] \left[ k^{1}-\left( -%
\mathcal{I}_{4}\right) ^{1/4}\tau \right] \left[ k^{2}-\left( -\mathcal{I}%
_{4}\right) ^{1/4}\tau \right]^{2}  -b^{2};  \notag \\
&&  \label{ugeneral} \\
\ x_{non-BPS,Z\neq 0}^{1}\left( \tau \right)
&=&\lambda_{0}^{-1}e^{\alpha
_{1}}\cdot  \notag \\
&&\cdot \frac{\left[ k^{2}-\left( -\mathcal{I}_{4}\right) ^{1/4}\tau \right]^{2} %
-\left[a-\left( -\mathcal{I}_{4}\right) ^{1/4}\tau \right] \left[ k^{1}-\left( -%
\mathcal{I}_{4}\right) ^{1/4}\tau \right] }{\left[ k^{2}-\left( -\mathcal{I}%
_{4}\right) ^{1/4}\tau \right]^{2} +\left[ a-\left( -\mathcal{I}_{4}\right) ^{1/4}\tau %
\right] \left[ k^{1}-\left( -\mathcal{I}_{4}\right) ^{1/4}\tau \right] -2{b}}%
;  \notag \\
&&  \label{xgeneral1} \\
\ x_{non-BPS,Z\neq 0}^{2}\left( \tau \right)
&=&\lambda_{0}^{-1}e^{\alpha
_{2}}\cdot  \notag \\
&&\cdot \frac{\left[ k^{1}-a\right]\left[ k^{2}-
\left( -\mathcal{I}_{4}\right) ^{1/4}\tau \right]}{\left[ k^{1}+a-2\left( -\mathcal{I}%
_{4}\right) ^{1/4}\tau \right]\left[ k^{2}-\left( -\mathcal{I}%
_{4}\right) ^{1/4}\tau \right]-2{b}}%
;  \notag \\
&&  \label{xgeneral2} \\
y_{non-BPS,Z\neq 0}^{1}\left( \tau \right) &=&2\lambda_{0}
^{-1}e^{\alpha
_{1}}\cdot  \notag \\
&&\cdot \frac{exp\left[ -2U_{non-BPS,Z\neq 0}\left( \tau \right) \right] }{%
\left[ k^{2}-\left( -\mathcal{I}_{4}\right) ^{1/4}\tau \right]^{2} +\left[ a-\left( -%
\mathcal{I}_{4}\right) ^{1/4}\tau \right] \left[ k^{1}-\left( -\mathcal{I}%
_{4}\right) ^{1/4}\tau \right] -2{b}};  \notag \\
&&  \label{ygeneral1} \\
y_{non-BPS,Z\neq 0}^{2}\left( \tau \right) &=&2\lambda_{0}
^{-1}e^{\alpha
_{2}}\cdot  \notag \\
&&\cdot \frac{exp\left[ -2U_{non-BPS,Z\neq 0}\left( \tau \right) \right] }{%
\left[ k^{1}+a-2\left( -\mathcal{I}_{4}\right) ^{1/4}\tau \right]
\left[
k^{2}-\left( -\mathcal{I}_{4}\right) ^{1/4}\tau \right]  -2{b}},  \notag \\
&&  \label{ygeneral2}
\end{eqnarray}
where
\begin{equation}
\lambda _{0}\equiv (p/q)^{1/3},  \label{b-2-bis}
\end{equation}
$a\in \mathbb{R}_{0}$, $b$, $k^{i}\in \mathbb{R}$ ($k^{i}$s cannot all
vanish), and the doublet of real constants $\alpha _{i}$ satisfies the
constraint
\begin{equation}
\sum_{i=1,2}\alpha _{i}=0.  \label{b-2-tris}
\end{equation}
It is worth pointing out that the $D0-D6$ configuration supports \textit{%
axion-free} non-BPS $Z\neq 0$ attractor flow(s); when considering the
\textit{near-horizon limit}, and thus the critical, charge-dependent values
of the moduli, this is consistent with the analysis performed in \cite
{TT,TT2,Ceresole}. 
An axion-free attractor flow solution of Eqs. (\ref{ugeneral})-(\ref{ygeneral2}) can be obtained e.g. by putting
\begin{equation}
k^{i}=a~\forall i=1,2
\end{equation} 
and it reads as follows:
\begin{eqnarray}
exp\left[ -4U_{non-BPS,Z\neq 0,axion-free}\left( \tau \right) \right] &=&%
\left[ a-\left( -\mathcal{I}_{4}\right) ^{1/4}\tau \right]
^{4}-b^{2};
\notag  \label{night!-1} \\
\ x_{non-BPS,Z\neq 0,axion-free}^{i}\left( \tau \right) &=&0;
\notag
\label{night!-2} \\
y_{non-BPS,Z\neq 0,axion-free}^{i}\left( \tau \right)
&=&\lambda_{0}^{-1} e^{\alpha _{i}}\sqrt{\frac{\left[ a-\left(
-\mathcal{I}_{4}\right) ^{1/4}\tau \right] ^{2}+b}{\left[ a-\left(
-\mathcal{I}_{4}\right) ^{1/4}\tau \right] ^{2}-b}}.
\label{night!-3}
\end{eqnarray}

The non-BPS $Z\neq 0$ fake superpotential of the first-order
formalism can be computed to have the following form in the $D0-D6$
configuration:
\begin{eqnarray}
\mathcal{W}_{non-BPS,Z\neq
0}(z,\overline{z},q,p)=\frac{1}{4}\frac{1}{\sqrt{-i(s-\overline{s})(t-\overline{t})^{2}}}\left[
\left| q^{1/3}+p^{1/3}e^{-\alpha _{1}}s\right|\left|
q^{1/3}+p^{1/3}e^{-\alpha _{2}}t\right| ^{2}\right] \cdot
\notag \\
\notag \\
\cdot\left[ 1+2\frac{\left( q^{2/3}-p^{2/3}e^{-2\alpha _{1}}\left|
s\right| ^{2}\right) \left( q^{2/3}-p^{2/3}e^{-2\alpha _{j}}\left|
t\right| ^{2}\right) -e^{-\alpha _{1}-\alpha
_{2}}q^{2/3}p^{2/3}(s-\overline{s})(t-\overline{t})}{\left|
q^{1/3}+p^{1/3}e^{-\alpha _{1}}s\right|
^{2}\,\left| q^{1/3}+p^{1/3}e^{-\alpha _{2}}t\right| ^{2}} \right.+\notag \\
\notag \\
\left.+\frac{\left( q^{2/3}-p^{2/3}e^{-2\alpha _{2}}\left| t\right|
^{2}\right)^{2} -e^{-2\alpha
_{2}}q^{2/3}p^{2/3}(t-\overline{t})^{2}}{\left|
q^{1/3}+p^{1/3}e^{-\alpha _{2}}t\right| ^{4}}\right]
.\hspace{5.5cm} \notag \\
 \label{D0D6fake}
\end{eqnarray}

The axion-free version of such a fake superpotential  reads as
follows:
\begin{eqnarray}
\mathcal{W}_{non-BPS,Z\neq 0,axion-free}(y,q,p)=\frac{1}{2^{3}\sqrt{2}}%
\frac{1}{\sqrt{y^{1}}y^{2}}\left[ \left| q^{1/3}-ip^{1/3}e^{-\alpha
_{1}}y^{1}\right|\left|
q^{1/3}-ip^{1/3}e^{-\alpha _{2}}y^{2}\right|^{2} \right] \cdot  \notag \\
\notag \\
\cdot \left[ 1+2\frac{\left[ q^{2/3}-p^{2/3}e^{-2\alpha _{1}}\left(
y^{1}\right) ^{2}\right] \left[ q^{2/3}-p^{2/3}e^{-2\alpha
_{2}}\left( y^{2}\right) ^{2}\right] +4e^{-\left(\alpha _{1}+\alpha
_{2}\right)}q^{2/3}p^{2/3}y^{1}y^{2}}{\left|
q^{1/3}-ip^{1/3}e^{-\alpha _{1}}y^{1}\right| ^{2}\,\left|
q^{1/3}-ip^{1/3}e^{-\alpha _{2}}y^{2}\right|
^{2}} \right.+\notag \\
\notag \\
\left.+\frac{\left[ q^{2/3}+p^{2/3}e^{-2\alpha _{2}}\left(
y^{2}\right) ^{2}\right]^{2} }{\left| q^{1/3}-ip^{1/3}e^{-\alpha
_{2}}y^{2}\right| ^{4}}\right].\hspace{8.5cm}
\label{D0D6fake-axion-free}
\end{eqnarray}

Now, by exploiting the \textit{first-order formalism} \cite{Fake-Refs} for $%
d=4$ extremal BHs \cite{Cer-Dal-1,ADOT-1} (see also \cite{FHM} and \cite
{Gnecchi-1}), one can compute the relevant BH parameters of the non-BPS $%
Z\neq 0$ attractor flow of $d=4$ $st^{2}$ model in the $D0-D6$ configuration,
starting from the expression of the non-BPS $Z\neq 0$ \textit{fake
superpotential} $\mathcal{W}_{non-BPS,Z\neq 0}$ given by Eq. (\ref{D0D6fake}%
).

Eqs.(\ref{D0D6fake}) and (\ref{d-1}) yield, after some algebra, the
following expression for the \textit{ADM mass}:
\begin{gather}  \label{D0D6ADM}
M_{ADM,nBPS,Z\neq 0}\left( z_{\infty },\overline{z}_{\infty
},\Gamma _{D0-D6}\right) =\frac{P}{2^{5/2}}\cdot \left[ \left(\left[
\left( \Lambda ^{1}\right) ^{-1}+B^{1}\right]
^{2}+1\right)^{1/2}\left[ \left(
\Lambda ^{2}\right) ^{-2}+\left(B^{2}\right) ^{2}+1\right]\right.\notag \\
\notag \\
\left.+\left(\left[ \left( \Lambda ^{1}\right) ^{-1}-B^{1}\right]
^{2}+1\right)^{1/2} \left( \left[ \left( \Lambda ^{2}\right)
^{-2}-\left(B^{2}\right) ^{2}-1\right]^{2}-4\left( \Lambda
^{2}\right) ^{-2}\right)^{1/2}\right]
,  \notag \\
\end{gather}
where the quantities
\begin{equation}
\Lambda ^{i}\equiv \lambda_{0} y_{\infty }^{i};~~B^{i}\equiv
\frac{x_{\infty }^{i}}{y_{\infty }^{i}},~~P\equiv p y_{\infty
}^{2}\sqrt{y_{\infty }^{1}},~~Q\equiv \frac{q}{y_{\infty
}^{2}\sqrt{y_{\infty }^{1}}} \label{quantity}
\end{equation}
were introduced, and, for simplicity's sake, the $\alpha _{i}$s were
chosen all to vanish (i.e. $\alpha _{i}=0~\forall i=1,2$). $P$ and $Q$
are the \textit{dressed charges}, \textit{i.e.} a sort of \textit{%
asymptotical redefinition} of the charges pertaining to $D6$ and $D0$
branes, respectively. On the other hand, $\Lambda ^{i}$ and $B^{i}$ are
usually named \textit{(asymptotical brane) fluxes} and $B$\textit{-fields},
respectively and the following condition is met (see Eq. (5.44) of \cite{GLS-1} for $stu$ case) :
\begin{equation}
\Lambda ^{1}\left[ 1+\left( B^{1}\right) ^{2}\right] -\left( \Lambda
^{1}\right) ^{-1}=\Lambda ^{2}\left[ 1+\left( B^{2}\right)
^{2}\right] -\left( \Lambda ^{2}\right) ^{-1}.
\end{equation}

As observed in \cite{GLS-1}, Eq. (\ref{a-2}) (along with the definitions
(\ref{quantity})) yields the \textit{marginal bound} \cite{Marginal-Refs} to be
\textit{saturated}, because $M_{ADM,non-BPS,Z\neq 0}$ is equal to the sum of
the \textit{ADM masses} of four $D6$-branes with appropriate fluxes (for
further detail, see the discussion in \cite{GLS-1}).

Concerning the \textit{(covariant) scalar charges}, they can be
straightforwardly computed by recalling Eqs. (\ref{D0D6fake}) and (\ref{d-2}%
), but their expressions are rather cumbersome.
By denoting $s\equiv x^{1}-iy^{1} $ and $t\equiv x^{2}-iy^{2}$, the
covariant scalar charges of axion and dilaton in the $D0-D6$
configuration respectively read
\begin{eqnarray}
\Sigma _{x^{1},non-BPS,Z\neq 0}\left( x_{\infty }^{i},y_{\infty
}^{i},\Gamma _{D0-D6}\right) =\frac{P\,x_{\infty
}^{1}}{2\sqrt{2}(y_{\infty}^{1})^{2}}\frac{\left( \Lambda
^{1}\right) ^{-1}}{\left(1+\left[\left(
\Lambda ^{1}\right) ^{-1}+B^{1}\right]^{2}\right)^{1/2}}\cdot \hspace{3cm}\notag \\
\notag \\
\cdot \Bigg[\frac{\left(B^{2}\right)^{2}}{B^{1}}\left( \Lambda
^{1}\right) ^{-2}+\frac{1}{B^{1}}\left(\left( \Lambda ^{2}\right)
^{-1}-\left( \Lambda ^{1}\right)
^{-1}\right)^{2}+3B^{1}\left(\left(B^{2}\right)^{2}+1\right)+
\Lambda ^{1}\left(\left(B^{1}\right)^{2}+1\right)\left(\left(B^{2}\right)^{2}+1\right)\notag\\
+3\left(B^{2}\right)^{2}\left( \Lambda ^{1}\right) ^{-1}-2\left(
\Lambda ^{2}\right) ^{-1}+3\left(
\Lambda ^{1}\right) ^{-1}  \Bigg]; \hspace{8.5cm}\\
\notag \\
\Sigma _{x^{2},non-BPS,Z\neq 0}\left( x_{\infty }^{i},y_{\infty
}^{i},\Gamma _{D0-D6}\right) =\frac{P\,x_{\infty
}^{2}}{2\sqrt{2}(y_{\infty}^{2})^{2}}\frac{\left[\left( \Lambda
^{1}\right)
^{-1}B^{1}+\left(B^{1}\right)^{2}+1\right]}{\left(1+\left[\left(
\Lambda ^{1}\right) ^{-1}+B^{1}\right]^{2}\right)^{1/2}} ;\hspace{3.2cm}\\
\notag \\
\Sigma _{y^{1},non-BPS,Z\neq 0}\left( x_{\infty }^{i},y_{\infty
}^{i},\Gamma _{D0-D6}\right) =-\frac{P}{4\sqrt{2}\,y_{\infty
}^{1}}\,\frac{1}{\left(1+\left[\left(
\Lambda ^{1}\right) ^{-1}+B^{1}\right]^{2}\right)^{3/2}}\cdot \hspace{3cm}\notag \\
\notag \\
\cdot \Bigg[ \left(\Lambda^{1} \right)^{-4}\left(\Lambda^{2}
\right)^{-2}+3\left(\Lambda^{1} \right)^{-3}\left(\Lambda^{2}
\right)^{-2}B^{1}+\left(\Lambda^{1} \right)^{-2}\left(\Lambda^{2}
\right)^{-1}\left[3\left(\Lambda^{2} \right)^{-1}\left(
\left(B^{1}\right)^{2}+1\right)-2\left(\Lambda^{1}
\right)^{-1}\right]+\hspace{.5cm}\notag \\
\notag \\
+3\left(\Lambda^{1}\right)^{-1}B^{1}\left(
\left(B^{1}\right)^{2}-1\right) \left(
\left(B^{2}\right)^{2}+1\right) +\left( \left(B^{1}\right)^{4}-
1\right)\left(
\left(B^{2}\right)^{2}+1\right)-\left(\Lambda^{1}\Lambda^{2}\right)^{-2}\cdot \hspace{1.7cm} \notag \\
\notag \\
\cdot \left(
\left(B^{1}\right)^{2}-1\right)+\left(\Lambda^{1}\right)^{-1}B^{1}\left[\left(\Lambda^{2}\right)^{-2}\left
(B^{1}\right)^{2}+\left(\Lambda^{1}\right)^{-2}\left(
\left(B^{2}\right)^{2}+1\right) +3\left(\Lambda^{2}\right)^{-2}
-4\left(\Lambda^{1}\Lambda^{2}\right)^{-1}\right]\Bigg]; \\
\notag \\
\Sigma _{y^{2},non-BPS,Z\neq 0}\left( x_{\infty }^{i},y_{\infty
}^{i},\Gamma _{D0-D6}\right) =-\frac{P}{4\sqrt{2}\,y_{\infty
}^{2}}\,\frac{1}{\left(1+\left[\left(
\Lambda ^{1}\right) ^{-1}+B^{1}\right]^{2}\right)^{1/2}}\cdot \hspace{3cm}\notag \\
\notag \\
\cdot\Bigg[ \left(\Lambda^{1} \right)^{-2}\left(\Lambda^{2}
\right)^{-2}+\left(\Lambda^{1} \right)^{-1}\left(\Lambda^{2}
\right)^{-2}B^{2}+\left(\Lambda^{1} \right)^{-1}B^{1}\left(
\left(B^{2}\right)^{2}-1\right) +\left( \left(B^{1}\right)^{2}+
1\right)\left( \left(B^{2}\right)^{2}-1\right)\Bigg] .\hspace{.5cm}  \notag \\
\end{eqnarray}

\subsubsection{\label{D0-D6}$D0-D6$ with equal $B$-field: }

The non-BPS $Z\neq 0$ attractor flow in the $D0-D6$ configuration with equal 
$B$ fields are given by:
\begin{eqnarray}
exp\left[ -4U_{non-BPS,Z\neq 0}\left( \tau \right) \right] &=&\left[
k-\left( -\mathcal{I}_{4}\right) ^{1/4}\tau \right]^{2} \left[ h-\left( -%
\mathcal{I}_{4}\right) ^{1/4}\tau \right]^{2} -b^{2};  \notag \\
&&  \label{ugen} \\
\ x_{non-BPS,Z\neq 0}^{i}\left( \tau \right) &=&\lambda
_{0}^{-1}e^{\alpha
_{i}}\cdot  \notag \\
&&\cdot \frac{\left( h-k \right) %
\left[h-\left( -\mathcal{I}_{4}\right) ^{1/4}\tau \right]  }{\left[ h-\left( -\mathcal{I}%
_{4}\right) ^{1/4}\tau \right] \left[ h+k-2\left(
-\mathcal{I}_{4}\right)
^{1/4}\tau \right]  -2{b}}%
;  \notag \\
&&  \label{xgenB} \\
y_{non-BPS,Z\neq 0}^{i}\left( \tau \right) &=&2\lambda
_{0}^{-1}e^{\alpha
_{i}}\cdot  \notag \\
&&\cdot \frac{exp\left[ -2U_{non-BPS,Z\neq 0}\left( \tau \right) \right] }{%
\left[ h-\left( -\mathcal{I}_{4}\right) ^{1/4}\tau \right] \left[
h+k-2\left( -\mathcal{I}_{4}\right) ^{1/4}\tau \right] -2{b}},  \notag \\
&&  \label{ygenB}
\end{eqnarray}

An axion-free attractor flow solution  can be obtained e.g.by putting
\begin{equation}
k=h
\end{equation}
and it reads as follows:
\begin{eqnarray}
exp\left[ -4U_{non-BPS,Z\neq 0,axion-free}\left( \tau \right) \right] &=&%
\left[ k-\left( -\mathcal{I}_{4}\right) ^{1/4}\tau \right]
^{4}-b^{2};
\notag  \label{night!-1} \\
\ x_{non-BPS,Z\neq 0,axion-free}^{i}\left( \tau \right) &=&0; \notag
\label{night!-2} \\
y_{non-BPS,Z\neq 0,axion-free}^{i}\left( \tau \right) &=&\lambda
_{0}^{-1}e^{\alpha _{i}}\sqrt{\frac{\left[k-\left(
-\mathcal{I}_{4}\right) ^{1/4}\tau \right] ^{2}+b}{\left[ k-\left(
-\mathcal{I}_{4}\right) ^{1/4}\tau \right] ^{2}-b}}.
\label{night!-3}
\end{eqnarray}

The non-BPS $Z\neq 0$ fake superpotential
$\mathcal{W}_{non-BPS,Z\neq 0}(z,\overline{z},q,p)$ has the same previous
form by considering this condition: $lim_{\tau \rightarrow
0^{-}}s=lim_{\tau \rightarrow 0^{-}}t$ .\\

In this case we have the following expression for the ADM mass:
\begin{gather}  \label{D0D6ADM}
M_{ADM,non-BPS,Z\neq 0}\left( z_{\infty },\overline{z}_{\infty
},\Gamma _{D0-D6}\right) =\frac{P}{2^{7/2}}\,\left[ \left[ \left(
\Lambda \right) ^{-1}+B\right] ^{2}+1\right]^{\frac{3}{2}} \cdot  \notag \\
\cdot \left\{ 1+3\frac{\left[ \left( \Lambda \right) ^{-2}-\left(
B\right) ^{2}-1\right]^{2}  +4\left( \Lambda \right) ^{-2}}{\left[
\left[ \left( \Lambda \right) ^{-1}+B\right] ^{2}+1\right]^{2}
 }\right\} ,  \notag \\
\end{gather}
where
\begin{equation}
\Lambda ^{1}\equiv\Lambda ^{2}\equiv \Lambda  \label{e-1}\\
\end{equation}

Here covariant scalar charges of axion and dilaton for the two moduli fields $s$ and $t$
coincide and they respectively read as follows:
\begin{eqnarray}
\Sigma _{x,non-BPS,Z\neq 0}\left( x_{\infty },y_{\infty },\Gamma
_{D0-D6}\right) &=&3\sqrt{2}\,P\,x_{\infty }\frac{\left( \Lambda
^{-1}B+B^{2}+1\right) }{\sqrt{\left( \Lambda ^{-1}+B\right) ^{2}+1}}; \\
&&  \notag \\
\Sigma _{y,non-BPS,Z\neq 0}\left( x_{\infty },y_{\infty },\Gamma
_{D0-D6}\right) &=&-\frac{3P\,y_{\infty }}{\sqrt{2}}\,\frac{\left[
\Lambda
^{-4}+\Lambda ^{-3}B+\Lambda ^{-1}B\left( B^{2}-1\right) +B^{4}-1\right] }{%
\sqrt{\left( \Lambda ^{-1}+B\right) ^{2}+1}}.  \notag \\
&&
\end{eqnarray}

\subsection{\label{D0-D2-D4-D6}$D0-D2-D4-D6$: }
\subsubsection{\label{U-Duality-Transf}$U$-Duality Transformations
along the Orbit $\mathcal{O}_{non-BPS,Z\neq 0}$}
We want to transform from the
configuration $D0-D6$ (corresponding to charges $\left(
q_{0},p^{0}\right) $, which we denote here $\left( q,p\right) $) to
the most general configuration $D0-D2-D4-D6$, corresponding to all
BH charges switched on: $\left( q_{0},q_{i},p^{i},p^{0}\right) $.  The parameters $a_{i},b_{i},c%
_{i},d_{i}$ of the $M_{i}$s dualizing from $D0-D6$ to $D0-D2-D4-D6$
must satisfy the following set of constraints (see for e.g. equation(\ref{constraintsD0-D6})):

\begin{eqnarray}
-q_{0} &=&-a_{1}(a_{2})^{2}q+b_{1}(b_{2})^{2}%
p;  \notag \\
&&  \notag \\
q_{1} &=&-c_{1}(a_{2})^{2}q+%
d_{1}(b_{2})^{2}p;  \notag \\
&&  \notag \\
q_{2} &=&-\frac{1}{2}c_{2}a_{1}a_{2}q+\frac{1}{2}%
b_{2}d_{1}d_{2}p;\notag \\
&&  \notag \\
p^{1} &=&-a_{1}(c_{2})^{2}q+%
b_{1}(d_{2})^{2}p; \notag \\
&&  \notag \\
p^{2} &=&-a_{2}c_{1}c_{2}q+%
b_{2}d_{1}d_{2}p; \notag \\
&&  \notag \\
p^{0} &=&-c_{1}(c_{2})^{2}q+d_{1}(d_{2})^{2}%
p,  \label{a-1}
\end{eqnarray}
The explicit form of the matrices $M_{i}$s under consideration (and
of their inverse) reads as follows:
\begin{eqnarray}
M_{i} &=&-\frac{sgn\left(\lambda \right) }{\sqrt{\left( \varsigma
_{i}+\varrho _{i}\right) \lambda }}\left(
\begin{array}{cc}
\varsigma _{i}\lambda & -\varrho _{i} \\
\lambda & 1
\end{array}
\right) \Leftrightarrow M_{i}^{-1}=-\frac{sgn\left( \lambda \right) }{\sqrt{%
\left( \varsigma _{i}+\varrho _{i}\right) \lambda }}\left(
\begin{array}{cc}
1 & \varrho _{i} \\
-\lambda & \varsigma _{i}\lambda
\end{array}
\right) ;  \label{b-1} \\
&&  \notag \\
&&  \notag \\
\lambda &\equiv &\left( \frac{p}{q}\right) ^{1/3}\left[ \frac{%
2p^{1}(p^{2})^{2}+p^{0}\left( \sqrt{-\mathcal{I}_{4}}-p^{\Lambda
}q_{\Lambda }\right) }{2p^{1}(p^{2})^{2}-p^{0}\left(
\sqrt{-\mathcal{I}_{4}}-p^{\Lambda
}q_{\Lambda }\right) }\right] ^{1/3}\in \mathbb{R};  \label{b-2ex} \\
&&  \notag \\
&&  \notag \\
\varsigma _{1} &\equiv &\frac{\sqrt{-\mathcal{I}_{4}}+p^{0 }q_{0
}-p^{1}q_{1}+p^{2}q_{2}}{2(p^{2})^{2}-2p^{0}q_{1}}; \label{b-3} \\
&&  \notag \\
&&  \notag \\
\varsigma _{2} &\equiv &\frac{\sqrt{-\mathcal{I}_{4}}+p^{0 }q_{0
}+p^{1}q_{1}}{2p^{1}p^{2}-p^{0}q_{2}};  \label{b-4} 
\end{eqnarray}

\begin{eqnarray}
\varrho _{1} &\equiv &\frac{\sqrt{-\mathcal{I}_{4}}-p^{0 }q_{0
}+p^{1}q_{1}-p^{2}q_{2}}{2(p^{2})^{2}-2p^{0}q_{1}}; \label{b-5} \\
&&  \notag \\
&&  \notag \\
\varrho _{2} &\equiv &\frac{\sqrt{-\mathcal{I}_{4}}-p^{0 }q_{0
}-p^{1}q_{1}}{2p^{1}p^{2}-p^{0}q_{2}}.  \label{b-6}
\end{eqnarray}
The above definitions (\ref{b-3})-(\ref{b-6}) together with (\ref{I4}) imply that
\begin{eqnarray}
\varsigma _{1}\varrho _{1}&=&-\frac{(q_{2})^{2}+4q_{0}p^{1}}{%
4(p^{2})^{2}-4p^{0}q_{1}};   \label{b-4'}\\
\varsigma _{2}\varrho _{2}&=&-\frac{q_{1}q_{2}+2q_{0}p^{2}}{%
2p^{1}p^{2}-p^{0}q_{2}}.  \label{b-5'}
\end{eqnarray}
   The $U$-duality transformation (\ref{boss-1})-(\ref{boss-3}) belonging
to the orbit $\mathcal{O}_{non-BPS,Z\neq 0}$, leaves $%
\mathcal{I}_{4}$ unchanged:
\begin{equation}
\mathcal{I}_{4},st^{2}\left( p,q \right)
=-(p^{0}q_{0}+p^{1}q_{1})^{2}+(2p^{1}p^{2}-p^{0}q_{2})(2p^{2}q_{0}+q_{1}q_{2})=-%
\left( pq\right) ^{2}=\mathcal{I}_{4}\left( \Gamma _{D0-D6}\right) .
\end{equation}

\subsubsection{\label{D0-D2-D4-D6}$D0-D2-D4-D6$: the Most General
Flow and Fake Superpotential}
Now, by performing the $U$-duality transformation (\ref{boss-1}-(\ref{boss-3}%
) (along with Eqs. (\ref{b-1})-(\ref{b-5'})) and using the most general
non-BPS $Z\neq 0$ attractor flow in the $D0-D6$ configuration given by Eqs. (%
\ref{ugeneral})-(\ref{ygeneral2}), it is a matter of long but straightforward
computations to determine the most general non-BPS $Z\neq 0$ attractor flow
in the most general configuration, namely in the $D0-D2-D4-D6$ one, in which
\textit{all} BH charges are switched on. It reads as follows
\footnote{%
In the particular case in which $b=0$, the expression of $exp\left[
-4U_{non-BPS,Z\neq 0}\left( \tau \right) \right] $ can be recast in the form
\begin{equation*}
exp\left[ -4U_{non-BPS,Z\neq 0}\left( \tau \right) \right] =-\mathcal{I}%
_{4}\left( \mathcal{H}\left( \tau \right) \right) ,
\end{equation*}
consistently with the results of \cite{K2-bis} and \cite{GLS-1}, and on the
same ground of (the first of) Eqs. (\ref{BPS-sol}) and (\ref
{non-BPS-Z=0-flow}), respectively holding for the $\frac{1}{2}$-BPS and
non-BPS $Z=0$ attractor flows.} (the moduli are here denoted as $\mathcal{Z}%
^{i}\equiv \mathcal{X}^{i}-i\mathcal{Y}^{i}$; $i\neq j\neq l$ and no sum on
repeated $i-$indices throughout):

\begin{eqnarray}
exp\left[ -4U_{non-BPS,Z\neq 0}\left( \tau \right) \right]
&=&h_{0}\left(
\tau \right) h_{1}\left( \tau \right) h_{2}^{2}\left( \tau \right)  -b^{2};  \label{sol} \\
&&  \notag \\
\ \mathcal{X}_{non-BPS,Z\neq 0}^{1}\left( \tau \right)
&=&\frac{\left\{
\begin{array}{l}
\varsigma _{1}e^{2\alpha _{1}}\nu ^{2}\left[ h_{2}^{2}\left( \tau
\right) +h_{0}\left( \tau \right) h_{1}\left( \tau \right)
+2b\right] + \\
\\
+e^{\alpha _{1}}\nu (\varsigma _{1}-\varrho _{1})\left[
h_{2}^{2}\left( \tau \right) -h_{0}\left( \tau \right) h_{1}\left(
\tau
\right) \right] + \\
\\
-\varrho _{1}\left[ h_{2}^{2}\left( \tau \right) +h_{0}\left( \tau
\right) h_{1}\left( \tau \right) -2b\right]
\end{array}
\right\} }{\left\{
\begin{array}{l}
e^{2\alpha _{1}}\nu ^{2}\left[ h_{2}^{2}\left( \tau \right) +h_{0}\left( \tau \right) h_{1}\left( \tau \right) +2b\right] + \\
\\
+2e^{\alpha _{1}}\nu \left[ h_{2}^{2}\left( \tau \right)-h_{0}\left( \tau \right) h_{1}\left( \tau \right) \right] + \\
\\
+h_{2}^{2}\left( \tau \right) +h_{0}\left( \tau \right) h_{1}\left(
\tau \right) -2b
\end{array}
\right\} };  \notag \\
&&  \label{sol-01}
\\
\ \mathcal{Y}_{non-BPS,Z\neq 0}^{1}\left( \tau \right) &=&\frac{2\nu
e^{\alpha _{1}}(\varsigma _{1}+\varrho _{1})exp\left[
-2U_{non-BPS,Z\neq 0}\left( \tau \right) \right] }{\left\{
\begin{array}{l}
e^{2\alpha _{1}}\nu ^{2}\left[ h_{2}^{2}\left( \tau \right) +h_{0}\left( \tau \right) h_{1}\left( \tau \right) +2b\right] + \\
\\
+2e^{\alpha _{1}}\nu \left[ h_{2}^{2}\left( \tau \right)  -h_{0}\left( \tau \right) h_{1}\left( \tau \right) \right] + \\
\\
+h_{2}^{2}\left( \tau \right) +h_{0}\left( \tau \right) h_{1}\left(
\tau \right) -2b
\end{array}
\right\} },  \notag \\
&&  \label{sol-02}
\\
\ \mathcal{X}_{non-BPS,Z\neq 0}^{2}\left( \tau \right)
&=&\frac{\left\{
\begin{array}{l}
\varsigma _{2}e^{2\alpha _{2}}\nu ^{2}\left[ h_{1}\left( \tau
\right) h_{2}\left( \tau \right) +h_{0}\left( \tau \right)
h_{2}\left( \tau \right)
+2b\right] + \\
\\
+e^{\alpha _{2}}\nu (\varsigma _{2}-\varrho _{2})\left[  h_{1}\left(
\tau \right) h_{2}\left( \tau \right) -h_{0}\left( \tau \right)
h_{2}\left( \tau
\right) \right] + \\
\\
-\varrho _{2}\left[  h_{1}\left( \tau \right) h_{2}\left( \tau
\right) +h_{0}\left( \tau \right) h_{2}\left( \tau \right)
-2b\right]
\end{array}
\right\} }{\left\{
\begin{array}{l}
e^{2\alpha _{2}}\nu ^{2}\left[  h_{1}\left( \tau \right) h_{2}\left(
\tau \right) +h_{0}\left
( \tau \right) h_{2}\left( \tau \right) +2b\right] + \\
\\
+2e^{\alpha _{2}}\nu \left[  h_{1}\left( \tau \right) h_{2}\left( \tau \right)-h_{0}\left( \tau \right) h_{2}\left( \tau \right) \right] + \\
\\
+ h_{1}\left( \tau \right) h_{2}\left( \tau \right) +h_{0}\left(
\tau \right) h_{2}\left( \tau \right) -2b
\end{array}
\right\} };  \notag \\
&&  \label{sol-03} 
\end{eqnarray}
\begin{eqnarray}
\mathcal{Y}_{non-BPS,Z\neq 0}^{2}\left( \tau \right) &=&\frac{2\nu
e^{\alpha _{2}}(\varsigma _{2}+\varrho _{2})exp\left[
-2U_{non-BPS,Z\neq 0}\left( \tau \right) \right] }{\left\{
\begin{array}{l}
e^{2\alpha _{2}}\nu ^{2}\left[ h_{1}\left( \tau \right) h_{2}\left(
\tau
\right) +h_{0}\left( \tau \right) h_{2}\left( \tau \right) +2b\right] + \\
\\
+2e^{\alpha _{2}}\nu \left[ h_{1}\left( \tau \right) h_{2}\left(
\tau
\right) -h_{0}\left( \tau \right) h_{2}\left( \tau \right) \right] + \\
\\
+h_{1}\left( \tau \right) h_{2}\left( \tau \right) +h_{0}\left( \tau
\right) h_{2}\left( \tau \right) -2b
\end{array}
\right\} },  \notag \\
&&  \label{sol-04}
\end{eqnarray}
where $\varsigma _{i}$ and $\varrho _{i}$ have been defined in Eqs. (\ref
{b-3}) - (\ref{b-6}), respectively. The constants $\alpha _{i}$ and $b$
have been introduced in Eqs. (\ref{ugeneral}) - (\ref{ygeneral2}). Furthermore, the
new quantities (see Eqs. (\ref{b-2'}),(\ref{b-2-bis}) and (\ref{b-2ex}) as well)
\begin{eqnarray}
\nu &\equiv &\frac{\lambda }{\lambda _{0}}=\left[
\frac{2p^{1}\left(p^{2}\right)^{2}+p^{0}\left(
\sqrt{-\mathcal{I}_{4}}-p^{\Lambda }q_{\Lambda }\right) }{%
2p^{1}\left(p^{2}\right)^{2}-p^{0}\left(
\sqrt{-\mathcal{I}_{4}}-p^{\Lambda }q_{\Lambda
}\right) }\right] ^{1/3}\in \mathbb{R;}  \label{sol-3} \\
&&  \notag \\
h_{\Lambda }\left( \tau \right) &\equiv &b_{\Lambda }-\left( -\mathcal{I}%
_{4}\right) ^{1/4}\tau ,  \label{sol-4}
\end{eqnarray}
where $b_{\Sigma }$ are real constants, have been defined.

By performing the \textit{near-horizon} (\textit{i.e.} $\tau
\rightarrow -\infty $) \textit{limit}, Eqs. (\ref{sol-01}) -
(\ref{sol-04}), respectively yield the following 
\textit{critical} values of the moduli (the
subscript \textit{``}$H$\textit{''} stands for
\textit{``}horizon\textit{'')}:
\begin{eqnarray}
&&\mathcal{X}_{non-BPS,Z\neq 0,H}^{i}\equiv lim_{\tau \rightarrow
-\infty }\ \mathcal{X}_{non-BPS,Z\neq 0}^{i}\left( \tau \right)
=\frac{\varsigma _{i}e^{2\alpha _{i}}\nu ^{2}-\varrho
_{i}}{e^{2\alpha _{i}}\nu ^{2}+1};\,\,\,\,\,\,\,\,\forall i=1,2.
\label{attrx} \\
&&  \notag \\
&&\mathcal{Y}_{nBPS,Z\neq 0,H}^{1}\equiv lim_{\tau \rightarrow
-\infty }\
\mathcal{Y}_{nBPS,Z\neq 0}^{1}\left( \tau \right) =\frac{1}{2}\frac{%
e^{\alpha _{1}}(\varsigma _{1}+\varrho _{1})\nu }{e^{2\alpha _{1}}\nu ^{2}+1}%
=\frac{\sqrt{-\mathcal{I}_{4}}e^{\alpha _{1}}\nu }{\left(
2\left(p^{2}\right)^{2}-2p^{0}q_{1}\right) \left( e^{2\alpha
_{1}}\nu
^{2}+1\right) },  \notag \\
&&  \label{attry}
\\
&&\mathcal{Y}_{nBPS,Z\neq 0,H}^{2}\equiv lim_{\tau \rightarrow
-\infty }\
\mathcal{Y}_{nBPS,Z\neq 0}^{2}\left( \tau \right) =\frac{1}{2}\frac{%
e^{\alpha _{2}}(\varsigma _{2}+\varrho _{2})\nu }{e^{2\alpha _{2}}\nu ^{2}+1}%
=\frac{\sqrt{-\mathcal{I}_{4}}e^{\alpha _{2}}\nu }{\left(
2p^{1}p^{2}-p^{0}q_{2}\right) \left( e^{2\alpha _{2}}\nu
^{2}+1\right) }.  \notag \\
&&  \label{attry}
\end{eqnarray}
It is worth pointing out that the $D0-D2-D4-D6$ configuration does \textit{%
not} support \textit{axion-free} non-BPS $Z\neq 0$ attractor flow(s); when
considering the \textit{near-horizon limit}, and thus the critical,
charge-dependent values of the moduli, this is consistent with the analysis
performed in \cite{TT,TT2,Ceresole}.The solution (\ref{sol})-(\ref{sol-04}) 
(along with the definitions (\ref
{sol-3}) and (\ref{sol-4})) generalizes the result of \cite{K2-bis} for the particular
case of $st^{2}$. As mentioned in Sect. 1, of \cite{K2-bis}, it was shown that,
within the (non-BPS $Z\neq 0$-supporting branches of the) $D2-D6$ (\textit{%
electric}) and $D0-D2-D4-D6$ configurations, in absence of (some of the) $B$%
\textit{-fields} the attractor flow solution can be obtained by replacing
the $Sp\left( 6,\mathbb{R}\right) $-covariant vector $\Gamma $ of charges
(defined by Eq. (\ref{Gamma})) with the $Sp\left( 6,\mathbb{R}\right) $%
-covariant vector $\mathcal{H}\left( \tau \right) $ of harmonic functions
(defined by Eqs. (\ref{H})-(\ref{constraints})) in the corresponding
critical, horizon solution.

For the $\frac{1}{2}$-BPS and non-BPS $Z=0$ attractor flows, respectively
treated in Sects. \ref{BPS-Flow} and \ref{Non-BPS-Z=0-Flow}, such a
procedure allows one to determine the most general attractor flow solution
starting from the corresponding critical, horizon solution.

On the other hand, for the non-BPS $Z\neq 0$ attractor flow such a procedure
fails in presence of non-vanishing $B$\textit{-fields}. In other words, it
can be shown that the completely general non-BPS $Z\neq 0$ attractor flow
solution (\ref{sol})-(\ref{sol-04}) is \textit{not} a solution of the \textit{%
would-be non-BPS }$Z\neq 0$\textit{\ stabilization Eqs.} (see the
treatments of \cite{AoB-book}, \cite{K2} and \cite{BFMY} for $stu$ BHs).

The issue concerning whether (in all non-BPS $Z\neq 0$-supporting
configurations) the \textit{actual} \textit{non-BPS }$Z\neq 0$\textit{\
stabilization Eqs.} (if any) admit a \textit{(}$\frac{1}{2}$\textit{%
-)BPS-like }reformulation in terms of a non-BPS $Z\neq 0$ \textit{fake
superpotential} (whose general form is given by Eq. (\ref{D0D2D4D6fake})
below) is open, and its investigation is left for future work.

Next, we can compute the non-BPS $Z\neq 0$ \textit{fake superpotential} of
the \textit{first-order formalism} in the $D0-D2-D4-D6$ configuration. To do
this, we apply the $U$-duality transformation (\ref{boss-1})-(\ref{boss-3})
(along with Eqs. (\ref{b-1})-(\ref{b-6})) to the expression of the non-BPS $%
Z\neq 0$ \textit{fake superpotential} in the $D0-D6$ configuration given by
Eq. (\ref{D0D6fake}), and, by noticing that $\mathcal{W}$ does \textit{not}
have any further covariance property under such a transformation, after some
algebra one achieves the following result:
\begin{eqnarray}
\mathcal{W}_{non-BPS,Z\neq 0}(s,\overline{s},t,\overline{t}
,p^{0},p^{1},p^{2},q_{0},q_{1},q_{2})= \hspace{7.5cm} \notag \\
\notag \\
=\frac{1}{4}\frac{\nu ^{3/2}\left( -\mathcal{I}_{4}\right) ^{1/4}}{\sqrt{%
(\varsigma _{1}+\varrho _{1})}(\varsigma _{2}+\varrho
_{2})}e^{K/2}\left[\left| \varsigma _{1}-s+(\varrho
_{1}+s)e^{-\alpha _{1}}\nu ^{-1}\right| \left| \varsigma
_{2}-t+(\varrho _{2}+t)e^{-\alpha
_{2}}\nu ^{-1}\right|^{2}\right] \cdot  \notag \\
\notag \\
\cdot \left( 1+2\frac{\left[ |\varsigma _{1}-s|^{2}-|\varrho
_{1}+s|^{2}e^{-2\alpha _{1}}\nu ^{-2}\right] \left[ |\varsigma
_{2}-t|^{2}-|\varrho _{2}+t|^{2}e^{-2\alpha _{2}}\nu ^{-2}\right]
}{\left| \varsigma _{1}-s +(\varrho _{1}+s)e^{-\alpha _{1}}\nu
^{-1}\right|
^{2}\,\left| \varsigma _{2}-t+(\varrho _{2}+t)e^{-\alpha _{2}}\nu ^{-1}\right| ^{2}}+\hspace{2.5cm}\right.  \notag \\
\notag \\
\qquad -\left. 2\frac{e^{-(\alpha _{1}+\alpha _{2})}\nu
^{-2}(\varsigma _{1}+\varrho _{1})(\varsigma _{2}+\varrho
_{2})(s-\overline{s})(t-\overline{t})}{\left| \varsigma
_{1}-s+(\varrho _{1}+s)e^{-\alpha _{1}}\nu ^{-1}\right| ^{2}\,\left|
\varsigma _{2}-t+(\varrho _{2}+t)e^{-\alpha _{2}}\nu
^{-1}\right| ^{2}}+\hspace{4cm}\right.  \notag \\
\notag \\
+\left.\frac{\left[ |\varsigma _{2}-t|^{2}-|\varrho
_{2}+t|^{2}e^{-2\alpha _{2}} \nu ^{-2}\right]^{2} }{\left| \varsigma
_{2}-t+(\varrho _{2}+t)e^{-\alpha _{2}}\nu ^{-1}\right|
^{4}}-\frac{e^{-2\alpha _{2}}\nu ^{-2}(\varsigma _{2}+\varrho
_{2})^{2}(t-\overline{t})^{2}}{\left| \varsigma _{2}-t+(\varrho
_{2}+t)e^{-\alpha _{2}}\nu ^{-1}\right| ^{4}}\right).\hspace{3cm}
\label{D0D2D4D6fake}
\end{eqnarray}
Consistently with the \textit{first-order formalism} \cite{Fake-Refs} for $%
d=4$ extremal BHs \cite{Cer-Dal-1,ADOT-1} (see also \cite{FHM} and \cite
{Gnecchi-1}), it is easy to check that the \textit{near-horizon} \textit{%
limit} of $\mathcal{W}_{non-BPS,Z\neq 0}^{2}$ yields the square root of $-%
\mathcal{I}_{4}$ (given by Eq. (\ref{I4})), or equivalently the square root
of the \textit{Cayley's hyperdeterminant }$Det\left( \Psi \right) $:
\begin{gather}
\mathcal{W}_{non-BPS,Z\neq 0,H}^{2}(\Gamma _{D0-D2-D4-D6})\equiv  \notag \\
\equiv lim_{\tau \rightarrow -\infty }\mathcal{W}_{non-BPS,Z\neq 0}^{2}(%
\mathcal{Z}\left( \tau \right) ,\overline{\mathcal{Z}}\left( \tau \right)
,\Gamma _{D0-D2-D4-D6})=  \notag \\
=\sqrt{-\mathcal{I}_{4}}=\sqrt{Det\left( \Psi \right) }=\frac{%
S_{BH,non-BPZ,Z\neq 0}(\Gamma _{D0-D2-D4-D6})}{\pi },
\end{gather}
where in the last step the Bekenstein-Hawking entropy-area formula \cite
{hawking2} was used.

Now, as done for the $D0-D6$ configuration in the previous Subsection, by
exploiting the \textit{first-order formalism} \cite{Fake-Refs} for $d=4$
extremal BHs \cite{Cer-Dal-1,ADOT-1} (see also \cite{FHM} and \cite
{Gnecchi-1}), one can compute the relevant BH parameters, such as the
\textit{ADM mass} (Eq. (\ref{d-1})) and the \textit{covariant scalar charges}
(Eq. (\ref{d-2})), starting from the \textit{fake superpotential} $\mathcal{W%
}_{non-BPS,Z\neq 0}$ given by Eq. (\ref{D0D2D4D6fake}). The computations are
long but straightforward, and they yield cumbersome results 
which we thus decide to omit here. We will explicitly
analyze some particular configurations in Sect. \ref{Analysis-Particular}.

However, it is easy to realize that Eq. (\ref{D0D2D4D6fake}) implies the
\textit{marginal bound} \cite{Marginal-Refs} to be \textit{saturated},
because (see Eq. (\ref{d-1}))
\begin{gather}
M_{ADM,non-BPS,Z\neq 0,}(\mathcal{Z}_{\infty },\overline{\mathcal{Z}}%
_{\infty },\Gamma _{D0-D2-D4-D6})=  \notag \\
=\mathcal{W}_{non-BPS,Z\neq 0}(\mathcal{Z}_{\infty },\overline{\mathcal{Z}}%
_{\infty },\Gamma _{D0-D2-D4-D6})\equiv  \notag \\
\equiv lim_{\tau \rightarrow 0^{-}}\mathcal{W}_{non-BPS,Z\neq 0}(\mathcal{Z}%
\left( \tau \right) ,\overline{\mathcal{Z}}\left( \tau \right) ,\Gamma
_{D0-D2-D4-D6})
\end{gather}
is equal to the sum of the \textit{ADM masses} of four $D6$-branes with
appropriate fluxes (for further detail on definition of such brane fluxes,
see the related discussion in \cite{GLS-1}). Thus, generalizing the related
results of \cite{GLS-1} and \cite{Cai-Pang}, it can be stated that 
\textit{marginal stability} holds for the most general non-BPS $Z\neq 0$
attractor flow of the $\mathcal{N}=2$, $d=4$ $st^{2}$ model.

\section{\label{Analysis-Particular}Analysis of Particular Configurations}
In this Section we analyze in depth some particularly simple configurations,
generalizing some results in literature \cite{K2-bis,Hotta, GLS-1,Cai-Pang}, but for 
the much less known case of $st^{2}$ model rather than the popular $stu$ model.

\subsection{\label{D0-D4}\textit{Magnetic} ($D0-D4$)}

The configuration $D0-D4$ (also named \textit{magnetic}) of the $stu$ model
has been previously treated in \cite{Hotta} and \cite{GLS-1}. In thie case of $st^{2}$ model,
the quantities of the $U$-duality transformation (\ref{boss-1})-(\ref{boss-3}%
) along $\mathcal{O}_{non-BPS,Z\neq 0}$ defined by Eqs. (\ref{b-2ex})-(\ref
{b-6}) undergo a major simplification :
\begin{equation}
\lambda =\lambda _{0};~~\varsigma _{1}=\varrho
_{1}=\frac{\sqrt{-\mathcal{I}_{4}}}{2\left(p^{2}\right)^{2}}=
\sqrt{\frac{-q_{0}p^{1}}{(p^{2})^{2}}};~~\varsigma _{2}=\varrho
_{2}=\frac{\sqrt{-\mathcal{I}_{4}}}{2p^{1}p^{2}}=
\sqrt{\frac{-q_{0}}{p^{1}}}.
\end{equation} 
Correspondingly, the non-BPS $Z\neq 0$ attractor flow
(\ref{sol})-(\ref {sol-04}) acquires the following form
(as above, the moduli are here denoted
as $\mathcal{Z}^{i}\equiv \mathcal{X}^{i}-i\mathcal{Y}^{i}$; $i\neq j$
and no sum on repeated $i-$indices throughout):

\begin{eqnarray}
&&exp\left[ -4U_{non-BPS,Z\neq 0}\left( \tau \right) \right]
=h_{0}\left(
\tau \right) h_{1}\left( \tau \right) h_{2}^{2}\left( \tau \right)-b^{2};  \label{electric-1} \\
&&  \notag \\
\ &&\mathcal{X}_{non-BPS,Z\neq 0}^{1}\left( \tau \right) =\varsigma
_{1}\cdot
\notag \\
&&  \notag \\
&&\cdot \frac{e^{2\alpha _{1}}\left[ h_{2}^{2}\left( \tau \right)
+h_{0}\left( \tau \right) h_{1}\left( \tau \right) +2b\right] -%
\left[ h_{2}^{2}\left( \tau \right) +h_{0}\left( \tau \right)
h_{1}\left( \tau \right) -2b\right] }{\left\{
\begin{array}{l}
e^{2\alpha _{1}}\left[ h_{2}^{2}\left( \tau \right) +h_{0}\left(
\tau \right) h_{1}\left( \tau \right) +2b\right] +2e^{\alpha
_{1}}\left[ h_{2}^{2}\left( \tau \right)-h_{0}\left(
\tau \right) h_{1}\left( \tau \right) \right] + \\
\\
+h_{2}^{2}\left( \tau \right) +h_{0}\left( \tau \right) h_{1}\left(
\tau \right) -2b
\end{array}
\right\} };  \notag \\
&&  \notag \\
&&  \label{electric-2} \\
&&  \notag \\
\ &&\mathcal{X}_{non-BPS,Z\neq 0}^{2}\left( \tau \right) =\varsigma
_{2}\cdot
\notag \\
&&  \notag \\
&&\cdot \frac{e^{2\alpha _{2}}\left[ h_{1}\left( \tau \right)
h_{2}\left(
\tau \right) +h_{0}\left( \tau \right) h_{2}\left( \tau \right) +2b\right] -%
\left[ h_{1}\left( \tau \right) h_{2}\left( \tau \right)
+h_{0}\left( \tau \right) h_{2}\left( \tau \right) -2b\right]
}{\left\{
\begin{array}{l}
e^{2\alpha _{2}}\left[ h_{1}\left( \tau \right) h_{2}\left( \tau
\right) +h_{0}\left( \tau \right) h_{2}\left( \tau \right)
+2b\right] +2e^{\alpha _{2}}\left[ h_{1}\left( \tau \right)
h_{2}\left( \tau \right) -h_{0}\left(
\tau \right) h_{2}\left( \tau \right) \right] + \\
\\
+h_{1}\left( \tau \right) h_{2}\left( \tau \right) +h_{0}\left( \tau
\right) h_{2}\left( \tau \right) -2b
\end{array}
\right\} };  \notag \\
&&  \notag \\
&&  \label{electric-2}
\end{eqnarray}
\begin{eqnarray}
&&\mathcal{Y}_{non-BPS,Z\neq 0}^{1}\left( \tau \right) =4\varsigma
_{1}\cdot
\notag \\
&&  \notag \\
&&\cdot \frac{e^{\alpha _{1}}exp\left[ -2U_{non-BPS,Z\neq 0}\left(
\tau \right) \right] }{\left\{
\begin{array}{l}
e^{2\alpha _{1}}\left[ h_{2}^{2}\left( \tau \right) +h_{0}\left(
\tau \right) h_{1}\left( \tau \right) +2b\right] +2e^{\alpha
_{1}}\left[ h_{2}^{2}\left( \tau \right)-h_{0}\left(
\tau \right) h_{1}\left( \tau \right) \right] + \\
\\
+h_{2}^{2}\left( \tau \right) +h_{0}\left( \tau \right) h_{1}\left(
\tau \right) -2b
\end{array}
\right\} }.  \notag \\
&&  \notag \\
&&  \label{electric-3}
\\
&&\mathcal{Y}_{non-BPS,Z\neq 0}^{2}\left( \tau \right) =4\varsigma
_{2}\cdot
\notag \\
&&  \notag \\
&&\cdot \frac{e^{\alpha _{2}}exp\left[ -2U_{non-BPS,Z\neq 0}\left(
\tau \right) \right] }{\left\{
\begin{array}{l}
e^{2\alpha _{2}}\left[ h_{1}\left( \tau \right) h_{2}\left( \tau
\right) +h_{0}\left( \tau \right) h_{2}\left( \tau \right)
+2b\right] +2e^{\alpha _{2}}\left[ h_{1}\left( \tau \right)
h_{2}\left( \tau \right) -h_{0}\left(
\tau \right) h_{2}\left( \tau \right) \right] + \\
\\
+h_{1}\left( \tau \right) h_{2}\left( \tau \right) +h_{0}\left( \tau
\right) h_{2}\left( \tau \right) -2b
\end{array}
\right\} }.  \notag \\
&&  \label{electric-4}
\end{eqnarray}
It is worth pointing out that the $D0-D4$ configuration supports \textit{%
axion-free} non-BPS $Z\neq 0$ attractor flow(s); when considering the
\textit{near-horizon limit}, and thus the critical, charge-dependent values
of the moduli, this is consistent with the analysis performed in \cite
{TT,TT2,Ceresole} for the $stu$ case. An \textit{axion-free} attractor flow 
solution of Eqs. (%
\ref{electric-1})-(\ref{electric-4}) can be obtained \textit{e.g.} by
putting
\begin{eqnarray}
\alpha _{i} &=&0~\forall i=1,2.  \label{e-1-tris} \\
b &=&0,  \label{e-1-tris-2}
\end{eqnarray}
and it reads as follows:
\begin{eqnarray}
&&exp\left[ -4U_{non-BPS,Z\neq 0,axion-free}\left( \tau \right)
\right] =h_{0}\left( \tau \right) h_{1}\left( \tau \right)
h_{2}^{2}\left( \tau \right)
;  \notag  \label{night!-4} \\
\ &&\mathcal{X}_{non-BPS,Z\neq 0,axion-free}^{i}\left( \tau \right)
=0,\,\,\,\,\,i=1,2.
\notag  \label{night!-5} \\
&&\mathcal{Y}_{non-BPS,Z\neq 0,axion-free}^{1}\left( \tau \right)
=\varsigma _{1}\sqrt{\frac{h_{0}\left( \tau \right) h_{1}\left( \tau
\right)}{h_{2}^{2}\left( \tau \right) }}.
\notag  \label{night!-6} \\
&&\mathcal{Y}_{non-BPS,Z\neq 0,axion-free}^{2}\left( \tau \right)
=\varsigma _{2}\sqrt{\frac{h_{0}\left( \tau \right)}{h_{1}\left(
\tau \right) }}.  \label{night!-6}
\end{eqnarray}
Furthermore, within the additional assumption (\ref{e-1-tris}), Eqs. (\ref
{electric-1})-(\ref{electric-4}) yield the particular solution of the one obtained in \cite{GLS-1} for the $stu$ case.

Always considering a framework in which the assumption
(\ref{e-1-tris}) holds true, Eqs. (\ref{D0D2D4D6fake}) yields that
the non-BPS $Z\neq 0$ \textit{fake superpotential} in the $D0-D4$
configuration has the following expression:
\begin{gather}
\left. \mathcal{W}_{non-BPS,Z\neq 0}\right| _{\alpha _{i}=0~\forall
i}\left( \mathcal{Z},\overline{\mathcal{Z}},\Gamma _{D0-D4}\right)
=e^{K/2}\cdot \left[
-q_{0}+p^{1}\left|t\right|^{2}+p^{2}(s\bar{t}+t\bar{s}) \right] .
\label{D0D4fake}
\end{gather}
The \textit{axion-free} version of such a \textit{fake superpotential} %
 reads as follows:
\begin{equation}
\mathcal{W}_{non-BPS,Z\neq 0,axion-free}\left( \mathcal{Y},\Gamma
_{D0-D4}\right) =\frac{1}{2\sqrt{2}}\frac{\left[ -q_{0}+p^{1}\left(\mathcal{Y}^{2}\right)^{2}%
+2p^{2}\mathcal{Y}^{1}\mathcal{Y}^{2}\right]
}{\sqrt{\mathcal{Y}^{1}}\mathcal{Y}^{2}}.
\label{D0D4fake-axion-free}
\end{equation}

The existence of a \textit{first-order formalism} in the non-BPS $Z\neq 0$%
-supporting (branch of the) $D0-D4$ configuration of the $st^{2}$ model, based
on the \textit{fake superpotential} given by Eq. (\ref{D0D4fake}), gives a
simple explanation of the \textit{integrability} of the equations of motion
of scalars, for the $st^{2}$ case.

Now, as done above for the $D0-D6$ and $D0-D2-D4-D6$ configurations, by
exploiting the \textit{first-order formalism} for $d=4$ extremal BHs, one
can compute the relevant BH parameters, such as the \textit{ADM mass} and
the \textit{covariant scalar charges}, starting from the \textit{fake
superpotential} $\left. \mathcal{W}_{non-BPS,Z\neq 0}\right| _{\alpha
_{i}=0~\forall i}$ given by Eq. (\ref{D0D4fake}).

Concerning the \textit{ADM mass}, by recalling Eq. (\ref{d-1}) and using Eq.
(\ref{D0D4fake}) one obtains an explicit expression which, after introducing
suitable \textit{dressed charges} (see Eq. (\ref{f-1})) and putting (see Eq.
(\ref{quantity}))
\begin{equation}
B^{1}=B^{2}=B,  \label{e-1-bis}
\end{equation}
is given by Eq. (4.6) of \cite{GLS-1} for the $stu$ model, which we report here for
completeness' sake but for the $st^{2}$ model:
\begin{gather}
\left. M_{ADM,non-BPS,\,Z\neq 0}\right| _{\alpha _{i}=0~\forall
i}\left( \mathcal{Z}_{\infty },\overline{\mathcal{Z}}_{\infty
},\Gamma
_{D0-D4}\right) =  \notag \\
=lim_{\tau \rightarrow 0^{-}}\left. \mathcal{W}_{non-BPS,Z\neq
0}\right|
_{\alpha _{i}=0~\forall i}\left( \mathcal{Z}\left( \tau \right) ,\overline{%
\mathcal{Z}}\left( \tau \right) ,\Gamma _{D0-D4}\right) =  \notag \\
=\frac{1}{2\sqrt{2}}\left[ \left| Q_{0}\right| +\left(
1+B^{2}\right) \left(P^{1}+2P^{2}\right)\right] ,  \label{i-1}
\end{gather}
where the \textit{dressed charges} are defined as follows (no summation on
repeated indices; notice the different definitions with respect to the $%
D0-D6 $ configuration, whose \textit{dressed charges} are given by Eq. (\ref
{quantity})):
\begin{equation}
Q_{0}\equiv \frac{q_{0}}{\sqrt{\mathcal{Y}_{\infty
}^{1}\mathcal{Y}_{\infty
}^{2}\mathcal{Y}_{\infty }^{2}}},~~P^{i}\equiv \frac{\sqrt{\mathcal{Y}%
_{\infty }^{1}\mathcal{Y}_{\infty }^{2}\mathcal{Y}_{\infty }^{2}}}{\mathcal{Y%
}_{\infty }^{i}}\,p^{i},\,\,\,\,\,\,i=1,2.  \label{f-1}
\end{equation}

 By recalling Eq. (\ref{d-2}) and using Eq. (\ref{D0D4fake}),one can
compute the \textit{covariant scalar charges} of the non-BPS $Z\neq
0$ attractor flow in the $D0-D4$ configuration.Within the simplifying assumptions 
(\ref{e-1-tris}) and (\ref{e-1-bis}), one obtains the following explicit
expressions ($i\neq j$, no sum on repeated indices):

\begin{gather}
\Sigma _{\mathcal{X},i,non-BPS,Z\neq 0}\left( \mathcal{X}_{\infty },\mathcal{%
Y}_{\infty },\Gamma _{D0-D4}\right) \equiv  \notag \\
\equiv lim_{\tau \rightarrow 0^{-}}\left( \frac{\partial \left. \mathcal{W}%
_{non-BPS,Z\neq 0}\right| _{\alpha _{m}=0~\forall m}}{\partial \mathcal{X}%
^{i}}\right) \left( \mathcal{Z}\left( \tau \right) ,\overline{\mathcal{Z}}%
\left( \tau \right) ,\Gamma _{D0-D4}\right);
\end{gather}
\begin{gather}
\Sigma _{\mathcal{Y},i,non-BPS,Z\neq 0}\left( \mathcal{X}_{\infty },\mathcal{%
Y}_{\infty },\Gamma _{D0-D4}\right) \equiv  \notag \\
\equiv lim_{\tau \rightarrow 0^{-}}\left( \frac{\partial \left. \mathcal{W}%
_{non-BPS,Z\neq 0}\right| _{\alpha _{m}=0~\forall m}}{\partial \mathcal{Y}%
^{i}}\right) \left( \mathcal{Z}\left( \tau \right) ,\overline{\mathcal{Z}}%
\left( \tau \right) ,\Gamma _{D0-D4}\right) ,
\end{gather}
then we can compute,
\begin{gather}
\Sigma _{\mathcal{X},1,non-BPS,Z\neq 0}\left( \mathcal{X}_{\infty },\mathcal{%
Y}_{\infty },\Gamma _{D0-D4}\right)=  \notag \\
=2\sqrt{2}\,\mathcal{X}_{\infty }\,P^{2} \notag \\
\\
\Sigma _{\mathcal{X},2,non-BPS,Z\neq 0}\left( \mathcal{X}_{\infty },\mathcal{%
Y}_{\infty },\Gamma _{D0-D4}\right)=  \notag \\
=\sqrt{2}\,\mathcal{X}_{\infty }\,(P^{1}+P^{2})  \notag \\
\\
\Sigma _{\mathcal{Y},1,non-BPS,Z\neq 0}\left( \mathcal{X}_{\infty },\mathcal{%
Y}_{\infty },\Gamma _{D0-D4}\right) \notag \\
=\frac{\mathcal{Y}_{\infty }}{\sqrt{2}}\Big( -\left| Q_{0}\right|
-2P^{1}+(1-B^{2})(P^{1}+2P^{2})\Big) \notag \\
\\
\Sigma _{\mathcal{Y},2,non-BPS,Z\neq 0}\left( \mathcal{X}_{\infty },\mathcal{%
Y}_{\infty },\Gamma _{D0-D4}\right) \notag \\
=\frac{\mathcal{Y}_{\infty }}{\sqrt{2}}\Big( -\left| Q_{0}\right|
-2P^{2}+(1-B^{2})(P^{1}+2P^{2})\Big)
\end{gather}
where the split in \textit{axionic scalar charges} $\Sigma
_{\mathcal{X},i}$ and \textit{dilatonic scalar charges} $\Sigma
_{\mathcal{Y},i}$ was performed.\smallskip

Recalling that $\mathcal{W}_{\frac{1}{2}-BPS}=\left| Z\right| $ in all $%
\mathcal{N}=2$, $d=4$ supergravities, and using Eqs. (\ref{g-1}), (\ref{d-1}%
) and (\ref{d-2}), for the $\frac{1}{2}$-BPS attractor flow one
obtains the following expressions :
\begin{gather}
M_{ADM,\frac{1}{2}-BPS}\left( \mathcal{Z}_{\infty },\overline{\mathcal{Z}}%
_{\infty },\Gamma _{D0-D4}\right) =lim_{\tau \rightarrow
0^{-}}\left| Z\right| \left( \mathcal{Z}\left( \tau \right)
,\overline{\mathcal{Z}}\left(
\tau \right) ,\Gamma _{D0-D4}\right) =  \notag \\
=\frac{1}{2\sqrt{2}}\sqrt{\left( 1+B^{2}\right) ^{2}\left(
P^{1}+2P^{2}\right) ^{2}-2\left( -1+B^{2}\right)
Q_{0}(P^{1}+2P^{2})+Q_{0}^{2}};  \label{h-1}
\end{gather}
\begin{gather}
\Sigma _{\mathcal{X},1,\frac{1}{2}-BPS}\left( \mathcal{X}_{\infty },\mathcal{%
Y}_{\infty },\Gamma _{D0-D4}\right) \equiv lim_{\tau \rightarrow
0^{-}}\left( \frac{\partial \left| Z\right| }{\partial \mathcal{X}}%
\right) \left( \mathcal{Z}\left( \tau \right)
,\overline{\mathcal{Z}}\left(
\tau \right) ,\Gamma _{D0-D4}\right) = \hspace{3cm} \notag \\
=\frac{\mathcal{X}_{\infty }}{\,M_{ADM,\frac{1}{2}-BPS}}\, P^{2}
\,\left[ \left( 1+B^{2}\right) (P^{1}+2P^{2})-Q_{0}\right]
;  \label{h-2} \notag \\
\\
\Sigma _{\mathcal{X},2,\frac{1}{2}-BPS}\left( \mathcal{X}_{\infty },\mathcal{%
Y}_{\infty },\Gamma _{D0-D4}\right) \equiv lim_{\tau \rightarrow
0^{-}}\left( \frac{\partial \left| Z\right| }{\partial \mathcal{X}}%
\right) \left( \mathcal{Z}\left( \tau \right)
,\overline{\mathcal{Z}}\left(
\tau \right) ,\Gamma _{D0-D4}\right) = \hspace{3cm} \notag \\
=\frac{\mathcal{X}_{\infty
}^{2}}{2\,M_{ADM,\frac{1}{2}-BPS}}\,\left( P^{1}+P^{2}\right)
\,\left[ \left( 1+B^{2}\right) (P^{1}+2P^{2})-Q_{0}\right] ;
\label{h-2}
\end{gather}
\begin{gather}
\Sigma _{\mathcal{Y},1,\frac{1}{2}-BPS}\left( \mathcal{X}_{\infty },\mathcal{%
Y}_{\infty },\Gamma _{D0-D4}\right) \equiv lim_{\tau \rightarrow
0^{-}}\left( \frac{\partial \left| Z\right| }{\partial \mathcal{Y}}%
\right) \left( \mathcal{Z}\left( \tau \right)
,\overline{\mathcal{Z}}\left(
\tau \right) ,\Gamma _{D0-D4}\right) = \hspace{3cm}  \notag \\
=-\frac{\mathcal{Y}_{\infty }}{4M_{ADM,\frac{1}{2}-BPS}}\cdot  \notag \\
\cdot \left[ Q_{0}^{2}+2\left( 1+B^{2}\right)
P^{1}(P^{1}+2P^{2})+2\left( P^{1}-B^{2}(P^{1}+2P^{2})\right)
Q_{0}+(B^{4}-1)\left(P^{1}+2P^{2}\right)
^{2}\right] .  \label{h-3} \notag \\
\\
\Sigma _{\mathcal{Y},2,\frac{1}{2}-BPS}\left( \mathcal{X}_{\infty },\mathcal{%
Y}_{\infty },\Gamma _{D0-D4}\right) \equiv lim_{\tau \rightarrow
0^{-}}\left( \frac{\partial \left| Z\right| }{\partial \mathcal{Y}}%
\right) \left( \mathcal{Z}\left( \tau \right)
,\overline{\mathcal{Z}}\left(
\tau \right) ,\Gamma _{D0-D4}\right) = \hspace{3cm} \notag \\
=-\frac{\mathcal{Y}_{\infty }}{4M_{ADM,\frac{1}{2}-BPS}}\cdot  \notag \\
\cdot \left[ Q_{0}^{2}+2\left( 1+B^{2}\right)
P^{2}(P^{1}+2P^{2})+2\left( P^{2}-B^{2}(P^{1}+2P^{2})\right)
Q_{0}+(B^{4}-1)\left(P^{1}+2P^{2}\right) ^{2}\right] .  \label{h-3}
\end{gather}

For what concerns the non-BPS $Z=0$ attractor flow in the $D0-D4$
configuration, by recalling the treatment of Sect.
\ref{Non-BPS-Z=0-Flow} and using Eqs. (\ref{d-1}) and (\ref{d-2}),
one formally obtains the same
expressions (\ref{h-1})-(\ref{h-3}) for the \textit{ADM mass} and \textit{%
covariant (axionic/dilatonic) scalar charges}, with the only
difference of the magnetic charges $p^{1},p^{2}$ being changed to
their absolute values. The definition (\ref{quantity}) of the
$B$\textit{-field}(\textit{s})
must change accordingly; for instance, in the case $p^{1}>0,p^{2}<0$, 
the (unique, in the assumption (\ref{e-1-bis})) $B$\textit{%
-field} must be defined as follows:
\begin{equation}
B\equiv \frac{\mathcal{X}_{\infty }^{1}}{\mathcal{Y}_{\infty }^{1}}=-\frac{%
\mathcal{X}_{\infty }^{2}}{\mathcal{Y}_{\infty }^{2}},  \label{o-1}
\end{equation}
where $\mathcal{X}_{\infty }^{2}<0$.

Thence, one can for example compare the \textit{ADM masses}. Taking
into account the above results, it makes sense to compare only the
\textit{ADM masses} pertaining to the $\frac{1}{2}$-BPS and non-BPS
$Z\neq 0$ flows. By introducing the \textit{gap} $\Delta $ between the
squared \textit{ADM masses} as follows:
\begin{equation}
\Delta \left( \mathcal{X}_{\infty },\mathcal{Y}_{\infty },\Gamma
\right)
\equiv M_{ADM,non-BPS,Z\neq 0}^{2}\left( \mathcal{X}_{\infty },\mathcal{Y}%
_{\infty },\Gamma \right) -M_{\frac{1}{2}-BPS}^{2}\left(
\mathcal{X}_{\infty },\mathcal{Y}_{\infty },\Gamma \right) ,
\label{gap}
\end{equation}
and using Eqs. (\ref{i-1}) and (\ref{h-1}), one achieves the
following result, holding for the $D0-D4$ configuration of the
$st^{2}$ model:
\begin{eqnarray}
\Delta \left( \mathcal{X}_{\infty },\mathcal{Y}_{\infty },\Gamma
_{D0-D4}\right) &\equiv &\left. M_{non-BPS,\,Z\neq 0}^{2}\right|
_{\alpha _{i}=0~\forall i}\left( \mathcal{X}_{\infty
},\mathcal{Y}_{\infty },\Gamma
_{D0-D4}\right) -M_{\frac{1}{2}-BPS}^{2}\left( \mathcal{X}_{\infty },%
\mathcal{Y}_{\infty },\Gamma _{D0-D4}\right) =  \notag \\
&=&\frac{1}{2}B^{2}\left| Q_{0}\right| (P^{1}+2P^{2})\geqslant 0.
\label{gibson-1}
\end{eqnarray}
This is in a sense the difference generalizing the \textit{BPS
bound} \cite{gibbons2} to the whole attractor flow (in the non-BPS $Z\neq 0$%
-supporting branch of the \textit{magnetic} charge configuration).

$\Delta $ is dilaton-dependent and strictly positive all along the non-BPS $%
Z\neq 0$ attractor flow. At the infinity, by using the \textit{dressed
charges} defined by Eq. (\ref{f-1}), the result given by Eq. (4.8) of \cite
{GLS-1} is recovered for the particular case of $stu$ model. 
Thus, the \textit{BPS bound} \cite{gibbons2} holds not
only at the BH event horizon ($r=r_{H}$), but actually (in a
dilaton-dependent way) all along the non-BPS $Z\neq 0$ attractor flow (%
\textit{i.e.} $\forall r\geqslant r_{H}$). \medskip

Of course, by relaxing the simplifying conditions (\ref{e-1-tris}) and/or (%
\ref{e-1-bis}), \textit{i.e.} by considering non-vanishing $\alpha _{i}$s
(constrained by Eq. (\ref{b-2-tris})) and/or different, $i$-indexed $B$%
\textit{-fields}, a much richer situation arises, but the main features of
the framework, outlined above, are left unchanged.

\subsection{\label{D2-D6}\textit{Electric} ($D2-D6$)}
The configuration $D2-D6$ (also named \textit{electric}) of the $stu$ model
has been previously treated in \cite{K2-bis} and \cite{Cai-Pang}. Here we do
the same exercise for the less known $st^{2}$ model.
Analogously to what happens in the $D0-D4$ (\textit{magnetic})
configuration, in this case the quantities of the $U$-duality transformation
(\ref{boss-1})-(\ref{boss-3}) along $\mathcal{O}_{non-BPS,Z\neq 0}$ defined
by Eqs. (\ref{b-2ex})-(\ref{b-6}) undergo a major simplification (the prime
denotes the charges in the considered configuration):

\begin{equation}
\lambda =-\lambda _{0};~~\varsigma _{1}=\varrho _{1}=-\sqrt{\frac{
q_{2}^{\prime 2}}{4p^{\prime 0}q_{1}^{\prime }}}; ~~\varsigma
_{2}=\varrho _{2}=-\sqrt{\frac{ q_{1}^{\prime }}{p^{\prime 0}}}.
\end{equation}
Correspondingly, the non-BPS $Z\neq 0$ attractor flow
(\ref{sol})-(\ref {sol-04}) acquires the following form (as above,
the moduli are here denoted as $\mathcal{Z}^{i}\equiv
\mathcal{X}^{i}-i\mathcal{Y}^{i}$):
\begin{eqnarray}
&&exp\left[ -4U_{non-BPS,Z\neq 0}\left( \tau \right) \right]
=h_{0}\left(
\tau \right) h_{1}\left( \tau \right) h_{2}^{2}\left( \tau \right)  -b^{2};  \label{magnetic-1} \\
&&  \notag \\
\ &&\mathcal{X}_{non-BPS,Z\neq 0}^{1}\left( \tau \right) =\varsigma
_{1}\cdot
\notag \\
&&  \notag \\
&&\cdot \frac{e^{2\alpha _{1}}\left[ h_{2}^{2}\left( \tau \right)
+h_{0}\left( \tau \right) h_{1}\left( \tau \right) +2b\right] -%
\left[ h_{2}^{2}\left( \tau \right)+h_{0}\left( \tau \right)
h_{1}\left( \tau \right) -2b\right] }{\left\{
\begin{array}{l}
e^{2\alpha _{1}}\left[ h_{2}^{2}\left( \tau \right) +h_{0}\left(
\tau \right) h_{1}\left( \tau \right) +2b\right] -2e^{\alpha
_{1}}\left[ h_{2}^{2}\left( \tau \right)-h_{0}\left(
\tau \right) h_{1}\left( \tau \right) \right] + \\
\\
+h_{2}^{2}\left( \tau \right) +h_{0}\left( \tau \right) h_{1}\left(
\tau \right) -2b
\end{array}
\right\} };  \notag \\
&&  \label{magnetic-2} \\
&&  \notag \\
\ &&\mathcal{X}_{non-BPS,Z\neq 0}^{2}\left( \tau \right) =\varsigma
_{2}\cdot
\notag \\
&&  \notag \\
&&\cdot \frac{e^{2\alpha _{2}}\left[ h_{1}\left( \tau \right)
h_{2}\left(
\tau \right) +h_{0}\left( \tau \right) h_{2}\left( \tau \right) +2b\right] -%
\left[ h_{1}\left( \tau \right) h_{2}\left( \tau \right)
+h_{0}\left( \tau \right) h_{2}\left( \tau \right) -2b\right]
}{\left\{
\begin{array}{l}
e^{2\alpha _{2}}\left[ h_{1}\left( \tau \right) h_{2}\left( \tau
\right) +h_{0}\left( \tau \right) h_{2}\left( \tau \right)
+2b\right] -2e^{\alpha _{2}}\left[ h_{1}\left( \tau \right)
h_{2}\left( \tau \right) -h_{0}\left(
\tau \right) h_{2}\left( \tau \right) \right] + \\
\\
+h_{1}\left( \tau \right) h_{2}\left( \tau \right) +h_{0}\left( \tau
\right) h_{2}\left( \tau \right) -2b
\end{array}
\right\} };  \notag \\
&&  \label{magnetic-2} \\
&&  \notag \\
&&\mathcal{Y}_{non-BPS,Z\neq 0}^{1}\left( \tau \right) =-4\varsigma
_{1}\cdot
\notag \\
&&  \notag \\
&&\cdot \frac{e^{\alpha _{1}}exp\left[ -2U_{non-BPS,Z\neq 0}\left(
\tau \right) \right] }{\left\{
\begin{array}{l}
e^{2\alpha _{1}}\left[ h_{2}^{2}\left( \tau \right) +h_{0}\left(
\tau \right) h_{1}\left( \tau \right) +2b\right] -2e^{\alpha
_{1}}\left[ h_{2}^{2}\left( \tau \right)  -h_{0}\left(
\tau \right) h_{1}\left( \tau \right) \right] + \\
\\
+h_{2}^{2}\left( \tau \right)+h_{0}\left( \tau \right) h_{1}\left(
\tau \right) -2b
\end{array}
\right\} }.  \notag \\
&&  \label{magnetic-3} 
\end{eqnarray}
\begin{eqnarray}
&&\mathcal{Y}_{non-BPS,Z\neq 0}^{2}\left( \tau \right) =-4\varsigma
_{2}\cdot
\notag \\
&&  \notag \\
&&\cdot \frac{e^{\alpha _{2}}exp\left[ -2U_{non-BPS,Z\neq 0}\left(
\tau \right) \right] }{\left\{
\begin{array}{l}
e^{2\alpha _{2}}\left[ h_{1}\left( \tau \right) h_{2}\left( \tau
\right) +h_{0}\left( \tau \right) h_{2}\left( \tau \right)
+2b\right] -2e^{\alpha _{2}}\left[ h_{1}\left( \tau \right)
h_{2}\left( \tau \right) -h_{0}\left(
\tau \right) h_{2}\left( \tau \right) \right] + \\
\\
+h_{1}\left( \tau \right) h_{2}\left( \tau \right) +h_{0}\left( \tau
\right) h_{2}\left( \tau \right) -2b
\end{array}
\right\} }.  \notag \\
&&  \label{magnetic-3}
\end{eqnarray}
It is worth pointing out that the $D2-D6$ configuration supports \textit{%
axion-free} non-BPS $Z\neq 0$ attractor flow(s); when considering the
\textit{near-horizon limit}, and thus the critical, charge-dependent values
of the moduli, this is consistent with the analysis performed in \cite
{TT,TT2,Ceresole} for the $stu$ case. An \textit{axion-free} attractor flow solution of Eqs. (%
\ref{magnetic-1})-(\ref{magnetic-3}) can be obtained \textit{e.g.} by
assuming the conditions given by Eqs. (\ref{e-1-tris}) and (\ref{e-1-tris-2}%
), and it reads as follows:
\begin{eqnarray}
&&exp\left[ -4U_{non-BPS,Z\neq 0,axion-free}\left( \tau \right)
\right] =h_{0}\left( \tau \right) h_{1}\left( \tau \right)
h_{2}^{2}\left( \tau \right)
;  \label{night!!-1} \\
\ &&\mathcal{X}_{non-BPS,Z\neq 0,axion-free}^{i}\left( \tau \right)
=0,\,\,\,\,\,\forall i=1,2.
\notag  \label{night!!-2} \\
&&\mathcal{Y}_{non-BPS,Z\neq 0,axion-free}^{1}\left( \tau \right) =-%
\varsigma _{1}\sqrt{\frac{ h_{2}^{2}\left( \tau \right)
}{h_{0}\left( \tau \right)h_{1}\left( \tau
\right) }}; \notag  \label{night!!-3} \\
&&\mathcal{Y}_{non-BPS,Z\neq 0,axion-free}^{2}\left( \tau \right) =-%
\varsigma _{2}\sqrt{\frac{h_{1}\left( \tau \right) }{h_{0}\left(
\tau \right) }}.  \label{night!!-3}
\end{eqnarray}
As done for the \textit{magnetic} configuration, in order to further
simplify Eqs. (\ref{magnetic-1})-(\ref{magnetic-3}) and (\ref{night!!-1})-(%
\ref{night!!-3}), one can consider the particular case constrained by Eq. (%
\ref{e-1-tris}). Within such an additional assumption, if we perform a similar 
kind of computation for the $stu$ model, the solution obtained
in \cite{Cai-Pang}, generalizing the one of \cite{K2-bis}, can be recovered
as pointed out in \cite{stuunveiled}.

Furthermore, within the simplifying assumption (\ref{e-1-tris}), Eq.
(\ref
{D0D2D4D6fake}) yields that the non-BPS $Z\neq 0$ \textit{fake superpotential%
} in the $D2-D6$ configuration has the following expression:
\begin{gather}
\left. \mathcal{W}_{non-BPS,Z\neq 0}\right| _{\alpha _{i}=0~\forall
i}\left( \mathcal{Z},\overline{\mathcal{Z}},\Gamma _{D2-D6}\right)
=e^{K/2}\left| s\right| \left| t\right|^{2} \cdot \left[ p^{\prime
}{}^{0}+\frac{q_{1}^{\prime }}{\left|t\right|^{2}}
+\frac{q_{2}^{\prime }}{2}\frac{\left( s
\overline{t}+t\overline{s}\right) }{\left| s\right| ^{2}\left|
t\right| ^{2}}\right] . \label{D2D6fake}
\end{gather}
The \textit{axion-free} version of such a \textit{fake
superpotential}  (%
\textit{e.g.} pertaining to the solution (\ref{night!!-1})-(\ref{night!!-3}%
)) reads as follows:
\begin{equation}
\left. \mathcal{W}_{non-BPS,Z\neq 0}\right| _{\alpha _{i}=0~\forall
i,~axion-free}\left( \mathcal{Z},\overline{\mathcal{Z}},\Gamma
_{D2-D6}\right) =\frac{1}{2\sqrt{2}}\sqrt{\mathcal{Y}^{1}}\mathcal{Y}^{2}%
\left[ p^{\prime }{}^{0}+\frac{q_{1}^{\prime }}{\left(\mathcal{Y}%
^{2}\right)^{2}}+\frac{q_{2}^{\prime
}}{\mathcal{Y}^{1}\mathcal{Y}^{2}}\right] .
\label{D2D6fake-axion-free}
\end{equation}
The existence of a \textit{first-order formalism} in the non-BPS $Z\neq 0$%
-supporting (branch of the) $D2-D6$ configuration of the $st^{2}$ model, based
on the \textit{fake superpotential} given by Eq. (\ref{D2D6fake}), gives an
explanation of the \textit{integrability} of the equations of motion of
scalars supported by the \textit{electric} configuration (see the treatment
of \cite{Cai-Pang} applicable for the $stu$ case).

Now, as done above for the $D0-D6$, $D0-D2-D4-D6$ and $D0-D4$
configurations, by exploiting the \textit{first-order formalism} for $d=4$
extremal BHs, one can compute the relevant BH parameters, such as the
\textit{ADM mass} and the \textit{covariant scalar charges}, starting from
the \textit{fake superpotential} $\left. \mathcal{W}_{non-BPS,Z\neq
0}\right| _{\alpha _{i}=0~\forall i}$ given by Eq. (\ref{D2D6fake}).

Concerning the \textit{ADM mass}, by recalling Eq. (\ref{d-1}) and
using Eqs. (\ref{D2D6fake}), (\ref{quantity}) and (\ref{e-1-bis}), one
obtains an
explicit expression which, after introducing suitable \textit{dressed charges%
} (see Eq. (\ref{l-1})), reads as follows:
\begin{gather}
\left. M_{ADM,non-BPS,\,Z\neq 0}\right| _{\alpha _{i}=0~\forall
i}\left( \mathcal{Z}_{\infty },\overline{\mathcal{Z}}_{\infty
},\Gamma
_{D2-D6}\right) =  \notag \\
=lim_{\tau \rightarrow 0^{-}}\left. \mathcal{W}_{non-BPS,Z\neq
0}\right|
_{\alpha _{i}=0~\forall i}\left( \mathcal{Z}\left( \tau \right) ,\overline{%
\mathcal{Z}}\left( \tau \right) ,\Gamma _{D2-D6}\right) =  \notag \\
=\frac{\sqrt{1+B^{2}}}{2\sqrt{2}}\left[ (1+B^{2})P^{\prime
}{}^{0}+\left(Q_{1}^{\prime }+Q_{2}^{\prime }\right)\right] ,
\label{m-1}
\end{gather}
where the \textit{dressed charges} are defined as follows (no summation on
repeated indices; notice the different definitions with respect to the $%
D0-D6 $ and $D0-D4$ configurations, whose \textit{dressed charges} are given
by Eqs. (\ref{quantity}) and (\ref{f-1}), respectively):
\begin{equation}
P^{\prime }{}^{0}\equiv p^{\prime }{}^{0}\sqrt{\mathcal{Y}_{\infty }^{1}%
\mathcal{Y}_{\infty }^{2}\mathcal{Y}_{\infty }^{2}},\quad
Q_{i}^{\prime
}\equiv \frac{\mathcal{Y}_{\infty }^{i}}{\sqrt{\mathcal{Y}_{\infty }^{1}%
\mathcal{Y}_{\infty }^{2}\mathcal{Y}_{\infty }^{2}}}\,q_{i}^{\prime
}. \label{l-1}
\end{equation}

By recalling Eq. (\ref{d-2}) and using Eq. (\ref{D2D6fake}), one can compute
the \textit{covariant scalar charges} of the non-BPS $Z\neq 0$ attractor
flow in the $D2-D6$ configuration. Within the simplifying assumptions (\ref
{e-1-tris}) and (\ref{e-1-bis}), one obtains the following explicit
expressions :
\begin{gather}
\Sigma _{\mathcal{X},i,non-BPS,Z\neq 0}\left( \mathcal{X}_{\infty },\mathcal{%
Y}_{\infty },\Gamma _{D2-D6}\right) \equiv  \notag \\
\equiv lim_{\tau \rightarrow 0^{-}}\left( \frac{\partial \left. \mathcal{W}%
_{non-BPS,Z\neq 0}\right| _{\alpha _{m}=0~\forall m}}{\partial \mathcal{X}%
^{i}}\right) \left( \mathcal{Z}\left( \tau \right) ,\overline{\mathcal{Z}}%
\left( \tau \right) ,\Gamma _{D2-D6}\right);
\end{gather}
\begin{gather}
\Sigma _{\mathcal{Y},i,non-BPS,Z\neq 0}\left( \mathcal{X}_{\infty },\mathcal{%
Y}_{\infty },\Gamma _{D2-D6}\right) \equiv  \notag \\
\equiv lim_{\tau \rightarrow 0^{-}}\left( \frac{\partial \left. \mathcal{W}%
_{non-BPS,Z\neq 0}\right| _{\alpha _{m}=0~\forall m}}{\partial \mathcal{Y}%
^{i}}\right) \left( \mathcal{Z}\left( \tau \right) ,\overline{\mathcal{Z}}%
\left( \tau \right) ,\Gamma _{D2-D6}\right),
\end{gather}
\begin{gather}
\Sigma _{\mathcal{X},1,non-BPS,Z\neq 0}\left( \mathcal{X}_{\infty },\mathcal{%
Y}_{\infty },\Gamma _{D2-D6}\right) =  \notag \\
=\sqrt{2}\,\frac{\mathcal{X}_{\infty }^{1}}{\sqrt{1+B^{2}}}\,\left[
(1+B^{2})P^{\prime }{}^{0}+Q_{1}^{\prime }\right] \notag \\
\\
\Sigma _{\mathcal{X},2,non-BPS,Z\neq 0}\left( \mathcal{X}_{\infty },\mathcal{%
Y}_{\infty },\Gamma _{D2-D6}\right) =  \notag \\
=\sqrt{2}\,\frac{\mathcal{X}_{\infty }^{2}}{\sqrt{1+B^{2}}}\,\left[
(1+B^{2})P^{\prime }{}^{0}+\frac{Q_{2}^{\prime }}{2}\right]
\end{gather}
\begin{gather}
\Sigma _{\mathcal{Y},1,non-BPS,Z\neq 0}\left( \mathcal{X}_{\infty },\mathcal{%
Y}_{\infty },\Gamma _{D2-D6}\right)= \notag \\
=\frac{\mathcal{Y}_{\infty }^{1}}{\sqrt{2}\,\sqrt{1+B^{2}}}\left[
\left( B^{4}-1\right) P^{\prime }{}^{0}-2Q_{1}^{\prime }+\left(
1+B^{2}\right)
^{2}\left(Q_{1}^{\prime }+Q_{2}^{\prime }\right)\right] \notag \\
\\
\Sigma _{\mathcal{Y},2,non-BPS,Z\neq 0}\left( \mathcal{X}_{\infty },\mathcal{%
Y}_{\infty },\Gamma _{D2-D6}\right)= \notag \\
=\frac{\mathcal{Y}_{\infty }^{2}}{\sqrt{2}\,\sqrt{1+B^{2}}}\left[
\left( B^{4}-1\right) P^{\prime }{}^{0}-Q_{2}^{\prime }+\left(
1+B^{2}\right) ^{2}\left(Q_{1}^{\prime }+Q_{2}^{\prime
}\right)\right]
\end{gather}
where, as for the $D0-D4$ configuration, the split in
\textit{axionic scalar
charges} $\Sigma _{\mathcal{X},i}$ and \textit{dilatonic scalar charges} $%
\Sigma _{\mathcal{Y},i}$ was performed.\smallskip \\
 Recalling once
again that $\mathcal{W}_{\frac{1}{2}-BPS}=\left| Z\right| $
in all $\mathcal{N}=2$, $d=4$ supergravities, and using Eqs. (\ref{g-1}), (%
\ref{d-1}) and (\ref{d-2}), for the $\frac{1}{2}$-BPS attractor flow
one obtains the following expressions :
\begin{gather}
M_{ADM,\frac{1}{2}-BPS}\left( \mathcal{Z}_{\infty },\overline{\mathcal{Z}}%
_{\infty },\Gamma _{D2-D6}\right) =  \notag \\
=lim_{\tau \rightarrow 0^{-}}\left| Z\right| \left(
\mathcal{Z}\left( \tau \right) ,\overline{\mathcal{Z}}\left( \tau
\right) ,\Gamma _{D2-D6}\right) =
\notag \\
=\frac{\sqrt{1+B^{2}}}{2\sqrt{2}}\sqrt{(1+B^{2})^{2}\left( P^{\prime
}{}^{0}\right) ^{2}-2(-1+B^{2})\left| P^{\prime }{}^{0}\right|
\left(Q_{1}^{\prime }+Q_{2}^{\prime }\right)+ \left(Q_{1}^{\prime
}+Q_{2}^{\prime }\right) ^{2}}; \label{n-1}
\end{gather}
\begin{gather}
\Sigma _{\mathcal{X},1,\frac{1}{2}-BPS}\left( \mathcal{X}_{\infty },\mathcal{%
Y}_{\infty },\Gamma _{D2-D6}\right) \equiv lim_{\tau \rightarrow
0^{-}}\left( \frac{\partial \left| Z\right| }{\partial \mathcal{X}^{1}}%
\right) \left( \mathcal{Z}\left( \tau \right)
,\overline{\mathcal{Z}}\left(
\tau \right) ,\Gamma _{D2-D6}\right) =  \notag \\
=\frac{\mathcal{X}_{\infty }^{1}}{2M_{ADM,\frac{1}{2}-BPS}}\,\cdot  \notag \\
\cdot \left\{ \left( 1+B^{2}\right) ^{2}\left( P^{\prime
}{}^{0}\right) ^{2}+Q_{1}^{\prime }\left(Q_{1}^{\prime
}+Q_{2}^{\prime }\right)+\left| P^{\prime }{}^{0}\right| \left[
\left( -3+B^{2}\right) Q_{1}^{\prime }+\left( 1+B^{2}\right)
\left(Q_{1}^{\prime }+Q_{2}^{\prime }\right)\right] \right\} ;
\label{n-2}
\end{gather}
\begin{gather}
\Sigma _{\mathcal{X},2,\frac{1}{2}-BPS}\left( \mathcal{X}_{\infty },\mathcal{%
Y}_{\infty },\Gamma _{D2-D6}\right) \equiv lim_{\tau \rightarrow
0^{-}}\left( \frac{\partial \left| Z\right| }{\partial \mathcal{X}^{2}}%
\right) \left( \mathcal{Z}\left( \tau \right)
,\overline{\mathcal{Z}}\left(
\tau \right) ,\Gamma _{D2-D6}\right) =  \notag \\
=\frac{\mathcal{X}_{\infty }^{2}}{2M_{ADM,\frac{1}{2}-BPS}}\,\cdot  \notag \\
\cdot \left\{ \left( 1+B^{2}\right) ^{2}\left( P^{\prime
}{}^{0}\right) ^{2}+\frac{Q_{2}^{\prime }\left(Q_{1}^{\prime
}+Q_{2}^{\prime }\right)}{2}+\left| P^{\prime }{}^{0}\right| \left[
\frac{\left( -3+B^{2}\right) Q_{2}^{\prime }}{2}+\left(
1+B^{2}\right) \left(Q_{1}^{\prime }+Q_{2}^{\prime }\right)\right]
\right\} ;  \label{n-2}
\end{gather}
\\
\begin{gather}
\Sigma _{\mathcal{Y},1,\frac{1}{2}-BPS}\left( \mathcal{X}_{\infty },\mathcal{%
Y}_{\infty },\Gamma _{D2-D6}\right) \equiv lim_{\tau \rightarrow
0^{-}}\left( \frac{\partial \left| Z\right| }{\partial \mathcal{Y}^{1}}%
\right) \left( \mathcal{Z}\left( \tau \right)
,\overline{\mathcal{Z}}\left(
\tau \right) ,\Gamma _{D2-D6}\right) =  \notag \\
=-\frac{\mathcal{Y}_{\infty }^{1}}{4M_{ADM,\frac{1}{2}-BPS}}\cdot  \notag \\
\cdot \left\{
\begin{array}{l}
\left( -1+B^{2}\right) \left( 1+B^{2}\right) ^{2}\left( P^{\prime
}{}^{0}\right) ^{2}+\left( 1+B^{2}\right) \left(Q_{1}^{\prime }+Q_{2}^{\prime }\right) ^{2}+ \\
\\
+2Q_{1}^{\prime }\left[ \left( -1+3B^{2}\right) \left| P^{\prime
}{}^{0}\right| -\left(Q_{1}^{\prime }+Q_{2}^{\prime }\right)\right]
-2B^{2}\left( 1+B^{2}\right) \left| P^{\prime }{}^{0}\right|
\left(Q_{1}^{\prime }+Q_{2}^{\prime }\right)
\end{array}
\right\} .  \label{n-3}
\end{gather}
\begin{gather}
\Sigma _{\mathcal{Y},2,\frac{1}{2}-BPS}\left( \mathcal{X}_{\infty },\mathcal{%
Y}_{\infty },\Gamma _{D2-D6}\right) \equiv lim_{\tau \rightarrow
0^{-}}\left( \frac{\partial \left| Z\right| }{\partial \mathcal{Y}^{2}}%
\right) \left( \mathcal{Z}\left( \tau \right)
,\overline{\mathcal{Z}}\left(
\tau \right) ,\Gamma _{D2-D6}\right) =  \notag \\
=-\frac{\mathcal{Y}_{\infty }^{2}}{4M_{ADM,\frac{1}{2}-BPS}}\cdot  \notag \\
\cdot \left\{
\begin{array}{l}
\left( -1+B^{2}\right) \left( 1+B^{2}\right) ^{2}\left( P^{\prime
}{}^{0}\right) ^{2}+\left( 1+B^{2}\right) \left(Q_{1}^{\prime }+Q_{2}^{\prime }\right) ^{2}+ \\
\\
+Q_{2}^{\prime }\left[ \left( -1+3B^{2}\right) \left| P^{\prime
}{}^{0}\right| -\left(Q_{1}^{\prime }+Q_{2}^{\prime }\right)\right]
-2B^{2}\left( 1+B^{2}\right) \left| P^{\prime }{}^{0}\right|
\left(Q_{1}^{\prime }+Q_{2}^{\prime }\right)
\end{array}
\right\} .  \label{n-3}
\end{gather}

Hence, one can, as done for the \textit{magnetic} configuration in Subsect. \ref{D0-D4},
also for the \textit{electric} configuration, compute the
difference between the squared non-BPS $Z\neq 0$ \textit{fake superpotential}
and the squared absolute value of the $\mathcal{N}=2$, $d=4$ central charge
along the non-BPS $Z\neq 0$ attractor flow and compare the \textit{ADM masses}.
After al dusts get settled, one achieves the following result, holding for the $D2-D6$ configuration of the $st^{2}$ model:
\begin{gather}
\Delta \left( \mathcal{X}_{\infty },\mathcal{Y}_{\infty },\Gamma
_{D2-D6}\right) \equiv  \notag \\
\equiv \left. M_{non-BPS,\,Z\neq 0}^{2}\right| _{\alpha
_{i}=0~\forall i}\left( \mathcal{X}_{\infty },\mathcal{Y}_{\infty
},\Gamma _{D2-D6}\right) -M_{ADM,\frac{1}{2}-BPS}\left(
\mathcal{X}_{\infty },\mathcal{Y}_{\infty
},\Gamma _{D2-D6}\right) =  \notag \\
=\frac{1}{2}\left( 1+B^{2}\right) B^{2}\left| P^{\prime
}{}^{0}\right| \left(Q_{1}^{\prime }+Q_{2}^{\prime }\right)\geqslant
0.  \label{gibson-2}
\end{gather}

Unlike what happens for the \textit{magnetic} configuration, for
\textit{electric} configuration $\Delta $ does depend also on axions,
nevertheless it is still strictly positive all along the non-BPS $Z\neq 0$
attractor flow. At infinity, by using the \textit{dressed charges} defined
by Eq. (\ref{l-1}), the following result is achieved:
\begin{equation}
\Delta \left( \mathcal{X}_{\infty },\mathcal{Y}_{\infty },\Gamma \right) =%
\frac{P^{\prime 0}}{2}(1+B^{2})(Q_{1}^{\prime }+Q_{2}^{\prime }).
\end{equation}
Thus, the \textit{BPS bound} \cite{gibbons2} holds not only at the BH event
horizon ($r=r_{H}$), but actually (in a scalar-dependent way) all along the
non-BPS $Z\neq 0$ attractor flow (\textit{i.e.} $\forall r\geqslant r_{H}$).
\medskip

Of course, by relaxing the simplifying conditions (\ref{e-1-tris}) and/or (%
\ref{e-1-bis}), \textit{i.e.} by considering non-vanishing $\alpha _{i}$s
(constrained by Eq. (\ref{b-2-tris})) and/or different, $i$-indexed $B$%
\textit{-fields}, a much richer situation arises, but the main features of
the framework, outlined above, are left unchanged.\medskip

By noticing that the $D0-D4$ (\textit{magnetic}) and $D2-D6$ (\textit{%
electric}) configurations are reciprocally \textit{dual} in $d=4$ and
recalling the treatment of Subsect. \ref{U-Duality-Transform}, it is worth
computing the matrices $M_{i,D0-D4\longrightarrow D2-D6}$ representing the $%
U $-duality transformation along the charge orbit $\mathcal{O}%
_{non-BPS,Z\neq 0}$ which connects (the non-BPS $Z\neq 0$-supporting \textit{%
branches} of) such two charge configurations. In order to do this, we
exploit the treatment given in Subsect. \ref{U-Duality-Transform}, by
performing the following steps:
\begin{equation}
\overset{\left( q_{0},p^{i}\right) }{D0-D4}\longrightarrow \overset{\left(
q,p\right) }{D0-D6}\longrightarrow \overset{\left( q_{i}^{\prime },p^{\prime
0}\right) }{D2-D6}.
\end{equation}

For the step $D0-D4\longrightarrow D0-D6$, we consider $M_{i}^{-1}$
given by Eq. (\ref{b-1}), along with the definitions
(\ref{b-2ex})-(\ref{b-6})
specified for the configuration $D0-D4$, obtaining $M_{i,D0-D4%
\longrightarrow D0-D6}^{-1}$. Then, for the the step
$D0-D6\longrightarrow D2-D6$, we take $M_{i}$ given by Eq.
(\ref{b-1}), along with the definitions
(\ref{b-2ex})-(\ref{b-6}) specified for the configuration $D2-D6$, obtaining $%
M_{i,D0-D6\longrightarrow D2-D6}$. Thus (also recall Eq. (\ref{boss-1})):
\begin{eqnarray}
\left( M_{1,D0-D4\longrightarrow D2-D6}\right) _{a^{\prime
}}^{~b^{\prime }} &=&\left( M_{1,D0-D6\longrightarrow D2-D6}\right)
_{a^{\prime }}^{~a}\left(
M_{1,D0-D4\longrightarrow D0-D6}^{-1}\right) _{a}^{~b^{\prime }}=  \notag \\
&&  \notag \\
&=&
\begin{pmatrix}
0 & -\left(-\frac{q_{0}p^{1}\,q_{2}^{\prime 2}}{%
4p^{\prime }{}^{0}q_{1}^{\prime }\,\left(p^{2}\right)^{2}}\right)^{1/4} \\
\left(-\frac{4p^{\prime }{}^{0}q_{1}^{\prime }\,\left(p^{2}\right)^{2}}{%
q_{0}p^{1}\,q_{2}^{\prime 2}}\right)^{1/4} & 0
\end{pmatrix}
.
\end{eqnarray}
\begin{eqnarray}
\left( M_{2,D0-D4\longrightarrow D2-D6}\right) _{a^{\prime
}}^{~b^{\prime }} &=&\left( M_{i,D0-D6\longrightarrow D2-D6}\right)
_{a^{\prime }}^{~a}\left(
M_{2,D0-D4\longrightarrow D0-D6}^{-1}\right) _{a}^{~b^{\prime }}=  \notag \\
&&  \notag \\
&=&
\begin{pmatrix}
0 & -\left(-\frac{q_{0}\,q_{1}^{\prime }}{%
p^{\prime }{}^{0}\,p^{1}}\right)^{1/4} \\
\left(-\frac{p^{\prime }{}^{0}\,p^{1}}{%
q_{0}\,q_{1}^{\prime }}\right)^{1/4} & 0
\end{pmatrix}
.
\end{eqnarray}
Consequently, by recalling Eqs. (\ref{boss-3}) and ((\ref{b-4'}) - (\ref{b-5'})) one
can
directly relate the non-BPS $Z\neq 0$ attractor flows $\mathcal{Z}%
_{non-BPS,Z\neq 0,D2-D6}^{i}\left( \tau \right) $ and $\mathcal{Z}%
_{non-BPS,Z\neq 0,D0-D4}^{i}\left( \tau \right) $ (respectively
given by Eqs. (\ref{magnetic-1})-(\ref{magnetic-3}) and
(\ref{electric-1})-(\ref
{electric-4}); recall that $\mathcal{Z}^{i}\left( \tau \right) =\mathcal{X}%
^{i}\left( \tau \right) -i\mathcal{Y}^{i}\left( \tau \right) $) by
the
following expression, explicitly showing the \textit{duality} between the $%
D0-D4$ (\textit{magnetic}) and $D2-D6$ (\textit{electric})
configurations in $d=4$:
\begin{equation}
\mathcal{Z}_{non-BPS,Z\neq 0,D2-D6}^{1}\left( \tau \right) =-\sqrt{\frac{%
|q_{0}|\,p^{1}\,(q\prime _{2})^{2}}{4p\prime ^{0}\,q\prime
_{1}\,(p^{2})^{2}}}\,\frac{1}{\mathcal{Z}_{non-BPS,Z\neq
0,D0-D4}^{1}\left( \tau \right) }.
\end{equation}
\begin{equation}
\mathcal{Z}_{non-BPS,Z\neq 0,D2-D6}^{2}\left( \tau \right) =-\sqrt{\frac{%
|q_{0}|\,q\prime _{1}}{p\prime
^{0}\,p^{1}}}\,\frac{1}{\mathcal{Z}_{non-BPS,Z\neq
0,D0-D4}^{2}\left( \tau \right) }.
\end{equation}

\subsection{\label{D0-D2-D4}$D0-D2-D4$}
The configuration $D0-D2-D4$ of the $stu$ model has been previously treated
in \cite{Cai-Pang}, provoking us to do the same analysis for the less known case of $st^{2}$ model. In this case, the quantities of the $U$-duality
transformation (\ref{boss-1})-(\ref{boss-3}) along $\mathcal{O}%
_{non-BPS,Z\neq 0}$ defined by Eqs. (\ref{b-2ex})-(\ref{b-6}) have the
following form :
\begin{eqnarray}
&& \lambda=\lambda _{0}; \notag \\
&& \varsigma _{1}=\frac{\sqrt{-\mathcal{I}_{4}}%
-p^{1}q_{1}+p^{2}q_{2}}{2(p^{2})^{2}};~~\varrho _{1}=\frac{\sqrt{-\mathcal{I}_{4}}%
+p^{1}q_{1}-p^{2}q_{2}}{2(p^{2})^{2}}\notag \\
&& \varsigma _{2}=\frac{\sqrt{-\mathcal{I}_{4}}%
+p^{1}q_{1}}{2p^{1}p^{2}};~~\varrho _{2}=\frac{\sqrt{-\mathcal{I}_{4}}%
-p^{1}q_{1}}{2p^{1}p^{2}}.
\end{eqnarray}

Within the additional assumption (\ref{e-1-tris}) (considered for
simplicity' sake), the non-BPS $Z\neq 0$ attractor flow (\ref{sol})-(\ref
{sol-04}) correspondingly acquires the following form (as above, the moduli
are here denoted as $\mathcal{Z}^{i}\equiv \mathcal{X}^{i}-i\mathcal{Y}^{i}$%
):

\begin{eqnarray}
&&exp\left[ -4U_{non-BPS,Z\neq 0}\left( \tau \right) \right]
=h_{0}\left(
\tau \right) h_{1}\left( \tau \right) h_{2}^{2}\left( \tau \right) -b^{2};  \label{D0D2D4-1} \\
&&  \notag \\
\ &&\mathcal{X}_{non-BPS,Z\neq 0}^{1}\left( \tau \right) =\frac{\sqrt{-%
\mathcal{I}_{4}}}{2(p^{2})^{2}}\frac{b}{h_{2}^{2}\left( \tau \right)
}+\frac{p^{2}q_{2}-p^{1}q_{1}}{2(p^{2})^{2}};
\label{D0D2D4-2} \\
&&  \notag \\
\ &&\mathcal{X}_{non-BPS,Z\neq 0}^{2}\left( \tau \right) =\frac{\sqrt{-%
\mathcal{I}_{4}}}{2p^{1}p^{2}}\frac{b}{h_{1}\left( \tau \right)
h_{2}\left( \tau \right)
}+\frac{p^{1}q_{1}-p^{2}q_{2}}{2p^{1}p^{2}};
\label{D0D2D4-2} \\
&&  \notag \\
&&\mathcal{Y}_{non-BPS,Z\neq 0}^{1}\left( \tau \right) =\frac{\sqrt{-%
\mathcal{I}_{4}}}{2(p^{2})^{2}}\frac{exp\left[ -2U_{non-BPS,Z\neq
0}\left( \tau \right) \right] }{h_{2}^{2}\left( \tau \right) };
\label{D0D2D4-2} \\
&&  \notag \\
&&\mathcal{Y}_{non-BPS,Z\neq 0}^{2}\left( \tau \right) =\frac{\sqrt{-%
\mathcal{I}_{4}}}{2p^{1}p^{2}}\frac{exp\left[ -2U_{non-BPS,Z\neq
0}\left( \tau \right) \right] }{h_{1}\left( \tau \right) h_{2}\left(
\tau \right) }.  \label{D0D2D4-3}
\end{eqnarray}

It is worth pointing out that, as the general case $D0-D2-D4-D6$ (see
Subsect. \ref{D0-D2-D4-D6}), the $D0-D2-D4$ configuration does \textit{not}
support \textit{axion-free} non-BPS $Z\neq 0$ attractor flow(s); when
considering the \textit{near-horizon limit}, and thus the critical,
charge-dependent values of the moduli, this is consistent with the analysis
performed in \cite{TT,TT2,Ceresole} for the $stu$ model.

Furthermore, always within the simplifying assumption (\ref{e-1-tris}), Eq. (%
\ref{D0D2D4D6fake}) yields that the non-BPS $Z\neq 0$ \textit{fake
superpotential} in the $D0-D2-D4$ configuration has the following
expression:
\begin{gather}
\left. \mathcal{W}_{non-BPS,Z\neq 0}\right| _{\alpha _{i}=0~\forall
i}\left(
\mathcal{Z},\overline{\mathcal{Z}},\Gamma _{D0-D2-D4}\right) =  \notag \\
=e^{K/2}\left[ -q_{0}-\frac{q_{1}}{2}\left( s+\overline{%
s}\right) -\frac{q_{2}}{2}\left( t+\overline{%
t}\right) +p^{1}|t|^{2}+p^{2}\left(
s\overline{t}+t\overline{s}\right)\right] . \label{D0D2D4fake}
\end{gather}

The existence of a \textit{first-order formalism} in the non-BPS $Z\neq 0$%
-supporting (branch of the) $D0-D2-D4$ configuration of the $st^{2}$ model,
based on the \textit{fake superpotential} given by Eq. (\ref{D0D2D4fake}),
gives an explanation of the \textit{integrability} of the equations of motion
of scalars supported by such a configuration (see the treatment of \cite
{Cai-Pang} applicable for the $stu$ case).

Now, by exploiting the \textit{first-order formalism} for $d=4$ extremal
BHs, one can compute the relevant BH parameters, such as the \textit{ADM mass%
} and the \textit{covariant scalar charges}, starting from the \textit{fake
superpotential} $\left. \mathcal{W}_{non-BPS,Z\neq 0}\right| _{\alpha
_{i}=0~\forall i}$ given by Eq. (\ref{D0D2D4fake}).

Concerning the \textit{ADM mass}, by recalling Eq. (\ref{d-1}) and
using Eq. (\ref{D0D2D4fake}) one obtains the following result:

\begin{gather}
\left. M_{ADM,non-BPS,\,Z\neq 0}\right| _{\alpha _{i}=0~\forall
i}\left( \mathcal{Z}_{\infty },\overline{\mathcal{Z}}_{\infty
},\Gamma
_{D0-D2-D4}\right) =  \notag \\
=lim_{\tau \rightarrow 0^{-}}\left. \mathcal{W}_{non-BPS,Z\neq
0}\right|
_{\alpha _{i}=0~\forall i}\left( \mathcal{Z}\left( \tau \right) ,\overline{%
\mathcal{Z}}\left( \tau \right) ,\Gamma _{D0-D2-D4}\right) =  \notag \\
=\frac{1}{2\sqrt{2}}\left[ \left| Q_{0}\right|
-(Q_{1}\,B_{1}+Q_{2}\,B_{2})+(P^{1}+2P^{2})+(P^{1}B_{2}^{2}+2P^{2}B_{1}B_{2})%
\right] ,  \label{r-1}
\end{gather}
where the \textit{dressed charges} are defined by Eqs. (\ref{f-1}) and (\ref
{l-1}).

By recalling Eq. (\ref{d-2}) and using Eq. (\ref{D0D2D4fake}), one can
compute the \textit{covariant scalar charges} of the non-BPS $Z\neq 0$
attractor flow in the $D0-D2-D4$ configuration within the assumption
(\ref{e-1-tris}), and then one obtains the following explicit expressions :
\begin{gather}
\Sigma _{\mathcal{X},i,non-BPS,Z\neq 0}\left( \mathcal{X}_{\infty },\mathcal{%
Y}_{\infty },\Gamma _{D0-D2-D4}\right) \equiv  \notag \\
\equiv lim_{\tau \rightarrow 0^{-}}\left( \frac{\partial \left. \mathcal{W}%
_{non-BPS,Z\neq 0}\right| _{\alpha _{m}=0~\forall m}}{\partial \mathcal{X}%
^{i}}\right) \left( \mathcal{Z}\left( \tau \right) ,\overline{\mathcal{Z}}%
\left( \tau \right) ,\Gamma _{D0-D2-D4}\right);  \label{r-2}
\end{gather}
\begin{gather}
\Sigma _{\mathcal{Y},i,non-BPS,Z\neq 0}\left( \mathcal{X}_{\infty },\mathcal{%
Y}_{\infty },\Gamma _{D0-D2-D4}\right) \equiv  \notag \\
\equiv lim_{\tau \rightarrow 0^{-}}\left( \frac{\partial \left. \mathcal{W}%
_{non-BPS,Z\neq 0}\right| _{\alpha _{m}=0~\forall m}}{\partial \mathcal{Y}%
^{i}}\right) \left( \mathcal{Z}\left( \tau \right) ,\overline{\mathcal{Z}}%
\left( \tau \right) ,\Gamma _{D0-D2-D4}\right),  \label{r-3}
\end{gather}
and we can write
\begin{gather}
\Sigma _{\mathcal{X},1,non-BPS,Z\neq 0}\left( \mathcal{X}_{\infty },\mathcal{%
Y}_{\infty },\Gamma _{D0-D2-D4}\right)= \notag \\
=\sqrt{2}\mathcal{Y}_{\infty }^{1}\,\left(2P^{2}B_{2}-Q_{1}\right) \notag \\
\\
\Sigma _{\mathcal{X},2,non-BPS,Z\neq 0}\left( \mathcal{X}_{\infty },\mathcal{%
Y}_{\infty },\Gamma _{D0-D2-D4}\right)=  \notag \\
=\sqrt{2}\mathcal{Y}_{\infty }^{2}\,(P^{1}B_{2}+P^{2}B_{1}-Q_{2})
\end{gather}
\begin{gather}
\Sigma _{\mathcal{Y},1,non-BPS,Z\neq 0}\left( \mathcal{X}_{\infty },\mathcal{%
Y}_{\infty },\Gamma _{D0-D2-D4}\right)= \notag \\
=\frac{\mathcal{Y}_{\infty }^{1}}{\sqrt{2}}\left[ -\left|
Q_{0}\right|
-2P^{1}+(Q_{1}\,B_{1}+Q_{2}\,B_{2})+(P^{1}+2P^{2})-(P^{1}B_{2}^{2}+2P^{2}B_{1}B_{2})\right] \notag \\
\\
\Sigma _{\mathcal{Y},2,non-BPS,Z\neq 0}\left( \mathcal{X}_{\infty },\mathcal{%
Y}_{\infty },\Gamma _{D0-D2-D4}\right)= \notag \\
=\frac{\mathcal{Y}_{\infty }^{2}}{\sqrt{2}}\left[ -\left|
Q_{0}\right|
-2P^{2}+(Q_{1}\,B_{1}+Q_{2}\,B_{2})+(P^{1}+2P^{2})-(P^{1}B_{2}^{2}+2P^{2}B_{1}B_{2})\right]
\end{gather}
where, as above, the split in \textit{axionic scalar charges} $\Sigma _{%
\mathcal{X},i}$ and \textit{dilatonic scalar charges} $\Sigma _{\mathcal{Y}%
,i}$ was performed, and the definition (\ref{quantity}) of $B$\textit{-fields}
was used.\smallskip

As done for the \textit{magnetic} and\textit{\ electric} configurations
(respectively in Subsects. \ref{D0-D4} and \ref{D2-D6}), also for $D0-D2-D4$
configuration the difference between the squared
non-BPS $Z\neq 0$ \textit{fake superpotential} and the squared absolute
value of the $\mathcal{N}=2$, $d=4$ central charge along the non-BPS $Z\neq
0 $ attractor flow can be computed giving the result that $\Delta $ is strictly positive 
all along the non-BPS $Z\neq 0$ attractor flow, due to the fact that $\mathcal{I}_{4}$ 
is strictly negative.

Thus, the \textit{BPS bound} \cite{gibbons2} is found to hold not only at
the BH event horizon ($r=r_{H}$), but actually (in a scalar-dependent way)
all along the non-BPS $Z\neq 0$ attractor flow (\textit{i.e.} $\forall
r\geqslant r_{H}$). \medskip

It is here worth pointing out that, by exploiting the procedure outlined in
Sect.\ref{Analysis-Particular}, the results 
for $\Delta$ computation can be related one to the others by performing suitable $U$%
-duality transformations. In such a way, one can also
compute $\Delta $ for the non-BPS $Z\neq 0$-supporting branch of the most
general (\textit{i.e.} $D0-D2-D4-D6$) BH charge configuration.

\section{\label{Conclusion}Conclusion}
In the present paper the analysis and solution of the equations of motion of
the scalar fields of the so-called $st^{2}$ model \cite{BMOS-1},
consisting of $\mathcal{N}=2$, $d=4$ \textit{ungauged} supergravity coupled
to $2$ Abelian vector multiplets whose complex scalars span the special
K\"{a}hler manifold $\frac{G}{H}=\left( \frac{SU\left( 1,1\right) }{U\left(
1\right) }\right) ^{2}$, has been performed in full detail, filling the gap in the so far, existing supergravity black hole literature. The obtained results complete a so far unknown analysis of the $3$ classes of \textit{non-degenerate}  attractor flows of the $st^{2}$model in its full generality. It is to be noted that all their features have been studied and compared in this report. 

Various comments, remarks, ideas for further developments along the lines of
research considered in the present paper are listed below.

\begin{itemize}
\item Since the $st^{2}$ model is  a consistent truncation of the much known $stu$ model, 
given any classical solution in the $st^{2}$ model, it can be regarded as a classical solution
in the $stu$ model if we limit ourselves to a special class of solution for which the two moduli are equal i.e. $u=t$ and moreover, as this is the case, we should be able to derive all properties of the classical solutions in the $st^{2}$ model using the corresponding properties of the $stu$ model and then choosing a special subset of solutions for which $u=t$. This crucial fact has been cross-checked against all the results obtained in this paper and matched with what has been obtained from the corresponding results in \cite{GLS-1,stuunveiled}, after a degeneracy choice of $u=t$ is made.

\item By exploiting the approach considered in Sect. 5 of \cite{BMOS-1},
the $st^{2}$ model can be consistently related to the so-called $stu$ and $t^{3}$
models, respectively with $3$ and $1$ \textit{complex} scalars. Such
a procedure enables one to extend all the results obtained for the $st^{2}$ case to other SUGRA 
models. (For results holding true for $stu$ case, see \cite{stuunveiled}). 
Furthermore, by performing the \textit{near-horizon}
(\textit{i.e.} $\tau \rightarrow -\infty $) \textit{limit} on the attractor
flow solutions, the corresponding attractor solution at the
event horizon of the extremal BH can be obtained. This is particularly relevant for the
non-BPS $Z\neq 0$ horizon attractor solutions, hitherto analytically known
(in a rather intricate form) only for the $t^{3}$ model, so far the only $%
\mathcal{N}=2$, $d=4$ supergravity model based on \textit{cubic} special
K\"{a}hler geometry whose Attractor Eqs. had been completely solved. In the
\textit{near-horizon} \textit{limit}, the results of the present paper yield
the non-BPS $Z\neq 0$ horizon attractor flow solutions for the $st^{2}$
model.

\item The $st^{2}$ model has been recently shown to be relevant for the
Special entangled Quantum systems, Freudenthal construction and the study of 
the group of stochastic local operations and classical communications (SLOCC)
\cite{VLevay} with one distinguished \textit{qubit} and two bosonic \textit{qubits}
in \textit{quantum information theory} and extremal stringy BHs \cite{duff1}
(see also \cite{Levay-Saniga-Vrana} for further recent developments). In the
seventh of Refs. \cite{duff1} the relation between quantum information
theory and the theory of extremal stringy BHs was studied within the $stu$
model, showing that the \textit{three-qubit interpretation} of
supersymmetric, $\frac{1}{2}$-BPS attractors can be extended also to include
non-supersymmetric, non-BPS (both $Z\neq 0$ and $Z=0$) ones, performing a
classification of the attractor solutions based on the charge codes of
\textit{quantum error correction}. However, only \textit{double-extremal }%
solutions, with \textit{constant}, \textit{non-dynamical} scalars \textit{%
all along} the attractor flow, were discussed therein. Thus, as also
observed in \cite{Cai-Pang}, it would be interesting to extend the analysis
of the seventh of Refs. \cite{duff1} using the full general non-BPS (both $%
Z\neq 0$ and $Z=0$) attractor flow solutions obtained in the present paper
and see whether $st^{2}$ model does also find its applicability in the 
\textit{quantum information theory}.

\item The existence of a \textit{first-order formalism} for the equations
of motion for the scalar fields (also named \textit{attractor flow Eqs.}) in
the background of an extremal BH \cite{Cer-Dal-1,ADOT-1} \textit{in principle%
} implies the \textit{integrability} of such equations, regardless of their
final form. It is particularly relevant for the non-BPS $Z\neq 0$
attractor flow, as pointed out in Subsects. \ref{D0-D4}-\ref{D0-D2-D4}. It
would be interesting to study the integrability of the equations of motion
of the scalars in presence of \textit{quantum} (\textit{perturbative} and/or
\textit{non-perturbative}) corrections to the considered $st^{2}$ model. For
instance, it would be interesting to study the attractor flow Eqs. for a
quantum corrected prepotential $f=st^{2}+i\lambda $ \cite{BFMS2}, with $\lambda \in \mathbb{R%
}$, which is the only correction which preserves the \textit{Peccei-Quinn axion shift
symmetry} and modifies the geometry of the scalar manifold (see \cite{BFMS1}
and Refs. therein). A tempting idea, inspired by the intriguing connection
between quantum information theory and extremal BHs mentioned at the
previous point, is to consider the quantum, axion-shift-consistent parameter
$\lambda $ as related to the \textit{quantum noise} of the system (see
\textit{e.g.} \cite{Quantum-Noise} and Refs. therein).

\item  As found in \cite{Ferrara-Gimon}, also observed in \cite{GLS-1}
for the $stu$ model and noticed in Sect. \ref{BPS-Flow}, for the $st^{2}$ case 
an immediate consequence of the general form of $\frac{1}{2}$-BPS attractor flow given 
by Eq. (\ref{BPS-sol}) is that $\Gamma _{\infty }$  for the $st^{2}$ model 
satisfies the $\frac{1}{2}$\textit{-BPS Attractor Eqs. }\cite{BPS-flow-1}. 
This determines a sort of \textit{``Attractor
Mechanism at spatial infinity'',} mapping the $4$ \textit{real} moduli $%
\left( x^{1},x^{2},y^{1},y^{2}\right) $ into the $6$ \textit{real%
} constants $\left( p_{\infty }^{1},p_{\infty }^{2},q_{1,\infty },q_{2,\infty }\right) $, arranged as $%
\Gamma _{\infty }$ and constrained by the $2$ \textit{real} conditions (\ref
{constraints}). As noticed in \cite{GLS-1}, for the $stu$ model, the absence of \textit{flat}
directions in the $\frac{1}{2}$-BPS attractor flow for the $st^{2}$ model (which is a general
feature of $\mathcal{N}=2$, $d=4$ \textit{ungauged} supergravity coupled to
Abelian vector multiplets, \textit{at least} as far as the metric of the
scalar manifold is strictly positive-definite $\forall \tau \in \mathbb{R}%
^{-}$ \cite{FGK}) is crucial for the validity of the expression (\ref
{BPS-sol}). As pointed out in Sect. \ref{Non-BPS-Z=0-Flow}, the same holds
for the non-BPS $Z=0$ case. Indeed, a consequence of the general form of
non-BPS $Z=0$ attractor flow given by Eq. (\ref{non-BPS-Z=0-flow}) is that $%
\Gamma _{\infty }$ satisfies the \textit{non-BPS }$\mathit{Z=0}$ \textit{%
Attractor Eqs. }(see \textit{e.g.} \cite{AoB-book} and \cite{BFMY}),
determining a sort of \textit{``Attractor Mechanism at spatial infinity''}.
Analogously to what happens in the $\frac{1}{2}$-BPS case, the absence of
\textit{flat} directions in the non-BPS $Z=0$ attractor flow (which is
\textit{not} a general feature of $\mathcal{N}=2$, $d=4$ \textit{ungauged}
supergravity coupled to Abelian vector multiplets, but however holds for the
$st^{2}$ model \cite{Ferrara-Marrani-1,ferrara4}) is crucial for the validity
of the expression (\ref{non-BPS-Z=0-flow}). Bearing in mind the crucial
differences among the non-BPS $Z\neq 0$ attractor flow and the $\frac{1}{2}$%
-BPS and non-BPS $Z=0$ attractor flows, (such as the presence of a $1$-dim.
real \textit{moduli space} $SO\left( 1,1\right)$ \textit{%
all along} the non-BPS $Z\neq 0$ attractor flow), it would be interesting to
investigate whether there exists any non-BPS $Z\neq 0$ \textit{``Attractor Mechanism 
at spatial infinity''}.
\end{itemize}

\acknowledgments
The author would like to thank Alessio Marrani for enlightening discussion on various 
issues pertaining to Attractor Mechanism. The author is grateful to LNF, Frascati for kind 
hospitality and support. The work of the author has been supported 
in part by Dipartimento di Scienze Fisiche of Federico II University and 
INFN section of Napoli. The author also wants to thank Giampiero Esposito and Ashoke Sen 
for a careful reading of the manuscript and for their suggestions that led to the improvement
of the presentation of this paper.

\end{document}